\documentclass[11pt]{article}
\usepackage{amsthm,amsmath,epsf,amssymb,enumerate,array,color}
\usepackage[all,matrix,arrow,import,arc,poly,color,ps,dvips,knot]{xy}

\definecolor{dgreen}{RGB}{34, 139, 34} \definecolor{webdkgreen}{rgb}{0,0.3,0} \definecolor{blueviolet}{RGB}{138,43,226}
\definecolor{brown}{rgb}{.6,0,0} \definecolor{dblue}{rgb}{0,0,.7} \definecolor{indigo}{RGB}{50,0,105} 

\usepackage[dvips, hyperindex=true,
colorlinks=true, linkcolor=dblue,
citecolor=indigo, pagecolor=dblue,
bookmarksopen=false, urlcolor=dblue,
linktocpage]{hyperref}

\newtheorem{thm}{Theorem}[section]        \newtheorem{lemma}[thm]{Lemma}	
\newtheorem{definition}[thm]{Definition} \newtheorem{prop}[thm]{Proposition}   \newtheorem{example}[thm]{Example}
\newtheorem{conj}[thm]{Conjecture}  		

\DeclareFontFamily{U}{rsf}{} \DeclareFontShape{U}{rsf}{m}{n}{  <5> <6> rsfs5 <7> <8> <9> rsfs7 <10-> rsfs10}{}
\DeclareMathAlphabet\Scr{U}{rsf}{m}{n} \DeclareMathAlphabet\mathbi{U}{cmr}{bx}{it}

\def\CY{Calabi-Yau}
\def\roof{\mbox{\tiny \mbox{$\!\vee$}}}	\def\comp{\mbox{\scriptsize \mbox{$\circ \,$}}}

\def\O{\mathcal{O}} \def\c#1{\mathcal{#1}}	\def\Coh{\mathfrak{Coh}}
\def\C{{\mathbb C}}\def\P{{\mathbb P}} \def\F{{\mathbb F}}
 \def\Z{{\mathbb Z}}
\def\D{\mathrm{D}} \def\R{\mathbf{R}}
\def\iso{\cong} \def\bar{\overline}
\def\H{\operatorname{H}}

\def\Hom{\operatorname{Hom}} \def\sHom{\operatorname{\Scr{H}\!\!\textit{om}}}	
\def\Ext{\operatorname{Ext}}      		\def\RHom{\R\!\operatorname{Hom}}

\def\Spec{\operatorname{Spec}}

	\def\SL{\operatorname{SL}}		
	\def\U{\operatorname{U{}}}	

\def\rk{\operatorname{rk}}	\def\dim{\operatorname{dim}}

\def\ch{\operatorname{\mathrm{ch}}}		\def\td{\operatorname{\mathrm{td}}}

\def\Ltensor{\mathbin{\overset{\mathbf L}\otimes}}
\def\ff#1#2{{\textstyle\dfrac{#1}{#2}}}
\def\ms#1{\mathsf{#1}}		\def\cal{\mathcal}
\def\poso#1{#1\save="x"!LD+<0pt,-0.5mm>;  "x"!RD+<0pt,-0.5mm>**\dir{.}\restore}
\def\Cone#1{\mathsf{C}\! \left( #1 \right)}
\def\ses#1#2#3{\xymatrix@1{0 \ar[r] & #1 \ar[r] & #2 \ar[r] & #3 \ar[r] & 0}}

\def\pplogo{\vbox{\kern-\headheight\kern -29pt
\halign{##&##\hfil\cr&{\ppnumber}\cr\rule{0pt}{2.5ex}&\ppdate\cr}}}
\makeatletter
\def\ps@firstpage{\ps@empty \def\@oddhead{\hss\pplogo}%
  \let\@evenhead\@oddhead 
}
\def\maketitle{\par
 \begingroup
 \def\thefootnote{\fnsymbol{footnote}}
 \def\@makefnmark{\hbox{$^{\@thefnmark}$\hss}}
 \if@twocolumn
 \twocolumn[\@maketitle]
 \else \newpage
 \global\@topnum\z@ \@maketitle \fi\thispagestyle{firstpage}\@thanks
 \endgroup
 \setcounter{footnote}{0}
 \let\maketitle\relax
 \let\@maketitle\relax
 \gdef\@thanks{}\gdef\@author{}\gdef\@title{}\let\thanks\relax}
\makeatother

\begin{document}
\setcounter{page}0
\def\ppnumber{\vbox{\baselineskip14pt
\hbox{hep-th/0602165}}}
\def\ppdate{} \date{}

\title{\bf \LARGE On the $\C^n/\Z_m$ fractional branes		\\[10mm]}
\author{{\bf Robert L.~Karp}\thanks{rlk at vt.edu	}	\\[2mm]
\normalsize  Department of Physics,  Virginia Tech\\
\normalsize Blacksburg, VA 24061 USA				}

{\hfuzz=10cm\maketitle}

\vskip 1cm

\begin{abstract}
\normalsize
\noindent
We construct several geometric representatives for the $\C^n/\Z_m$ fractional branes on either a partially or the completely resolved orbifold. In the process we use large radius and conifold-type monodromies, and provide a strong consistency check. In particular, for  $\C^3/\Z_5$ we give three different sets of geometric representatives. We also find the explicit Seiberg-duality which connects our fractional branes to the ones given by the McKay correspondence. 
\end{abstract}

\vfil\break

\tableofcontents

\section{Introduction}    \label{s:intro}

In recent years there have been significant advances in our understanding of the physical properties of D-branes  throughout the entire moduli space of a given Calabi-Yau compactification. In particular, for Type II compactifications Douglas showed that the topological B-branes are in one-to-one correspondence with the objects of the derived category of coherent sheaves on the \CY\ variety \cite{Douglas:2000gi}. To relate the topological B-branes to the physical ones he also pioneered the notion of $\pi$-stability. Subsequently Bridgeland axiomatized $\pi$-stability.

Since the Kahler deformations are exact in the topological B-model, the derived category description of topological B-branes is valid at any point of the moduli space. On the other hand, $\c N=2$ Type II compactifications generically have a rich phase structure and the description of B-branes in the various phases is quite different. This gives rise to interesting mathematical statements, the best known of which is the celebrated McKay correspondence.

In general, determining the set of $\pi$-stable branes is cumbersome. One could start at a point with a good understanding of stability, e.g., a large radius point, where $\pi$-stability reduces to $\mu$-stability, and try to catalog what objects are lost and gained as the Kahler moduli are varied \cite{Aspinwall:2001dz}. 

Among the stable branes some of the most intriguing ones are the fractional branes. They are fractional in the following sense: assume that we are given a \CY\ 3-fold which develops a singularity somewhere in the Kahler moduli space. We can probe the singularity using a space-filling D3-brane. At this point in Kahler moduli space the probe D3-brane becomes (marginally) unstable and {\em decays} into the fractional branes. This picture applies in a rather broad context,  and is not restricted to orbifolds, although it was first discovered for orbifolds.

On the other hand, orbifolds provide a rich testing ground. In this case D-branes  can be described explicitly as boundary states in a solvable conformal field theory (CFT). The world-volume theory of the probe D3-brane, which is a quiver gauge theory \cite{Douglas:Moore}, gives a very different description of the D-branes, as objects in the derived category of representations of the  quiver. The McKay correspondence gives an equivalence between this category and the derived category of coherent sheaves on the resolved space \cite{Mukai:McKay}, as required by the topological string argument.

The McKay correspondence is a prototype of what happens in general: in different patches of the moduli space one has very different looking descriptions for the D-branes, which sit in {\em inequivalent} categories, but if one passes to the derived category then they all become equivalent. Therefore it makes sense to talk about a geometric representative for a brane at any point in moduli space. Passing from an abelian  category to the derived category is physically motivated by brane--anti-brane annihilation, thorough  tachyon condensation \cite{Douglas:2000gi,Witten:1998cd}. 

In the quiver language the  fractional branes are the simple representations, i.e., those that have no non-trivial subrepresentations. Although the  fractional branes are obvious in the quiver language, their geometric incarnation is unclear. The McKay correspondence tells us that there should be objects (bundles or perhaps complexes) on the resolved space whose $\Ext^1$-quiver is the one we started with. One of the central problems in this area is to find these objects. 

As a warm-up exercise one can try to determine the K-theory class of a fractional brane. So far, even this question has been answered only in a limited context, using mirror symmetry techniques \cite{Diaconescu:1999dt} or the McKay correspondence\footnote{ For the ample physics literature on this subject see, e.g., \cite{Paul:TASI2003} and references therein.}. 

A first goal of this paper is to get a deeper understanding of the geometry of  fractional branes, going beyond K-theory. We do this without resorting to mirror symmetry or the McKay correspondence. Instead we use the quantum symmetry of an orbifold theory to generate the fractional branes as an orbit.  This method has been successful for the $\C^2/\Z_n$ orbifolds \cite{en:fracC2}. 

Ultimately one would like to understand the world-volume theory of D-branes at an arbitrary point in the moduli space of a compact \CY . Studying examples where one has at his disposal different methods hopefully will teach us the ``mechanics'' of the geometric approach, and would certainly lead to results of phenomenological interest.

A second goal is to develop techniques to study monodromies in the Kahler moduli space. Monodromies played a crucial role in Seiberg-Witten theory \cite{SW:I}, and represent a subject of interest in itself. Using geometric engineering one can relate the Seiberg-Witten monodromies to D-brane monodromies \cite{en:Paul,en:Cinci}. As we will see shortly, the functors implementing the monodromy transformations are not simple by any measure. The relations they satisfy would be very hard to guess without physical input. By the end, we will have performed some very strong consistency checks. It is not surprising that a simple version of our functors, the Seidel-Thomas twist functors, plays an important role in Bridgeland's work on $\pi$-stability \cite{Bridgeland:quiver}.

A third motivation is to investigate what, if anything, can be gained by using stacks. Our investigations are ultimately very different from the ones pioneered by Sharpe and collaborators. Our approach relies on an extension of the McKay correspondence due to Kawamata, and naturally associates branes to regions of the moduli space where we have no solvable CFT description or reliable supergravity approximation. Fortunately, the algebro-geometric tools are powerful enough to produce several collections of fractional branes. This approach provides a detailed understanding of the $Y^{p,p}$ spaces \cite{Benvenuti}, as assumed in \cite{Herzog:2005sy}. It came as a surprise to the present author to realize that using stacks one can construct several collections of fractional branes for $\C^n/\Z_m$, for {\em any} $n$ and $m$! The same statement is far from being true without the use of  stacks.

Let us illustrate this point using the main workhorse of the paper, $\C^3/\Z_5$. This singularity has two partial resolutions, with exceptional divisors $\P^2(2,2,1)$ and $\P^2(1,1,3)$.\footnote{As a scheme $\P^2(2,2,1)$ is just $\P^2$, but as stacks they are different.} Viewing both of them as stacks, one has length five exceptional collections $\O,\O(1),\ldots ,\O(4)$. The candidate fractional branes are the push-forwards of the dual collections. We prove that both $\P^2(2,2,1)$ and $\P^2(1,1,3)$ lead to the same $\Ext^1$'s quiver.  Naturally, the details for the two cases are quite different, but the final result is the same. The superpotential comes from the $\Ext^2$'s and is related to an $A_\infty$-structure. So going from $\P^2(2,2,1)$ to $\P^2(1,1,3)$ does not change the quiver, but reinterprets the different terms in the superpotential in a non-trivial way.

Another motivation to understand fractional branes comes from the fact that they are related to each other through partial resolutions of the singularity.\footnote{I learned about this line of reasoning from Paul Aspinwall.}  For example, the resolution of the $\C^2/\Z_n$ singularity has an exceptional divisor which is reducible, with $n-1$ irreducible components $C_i=\P^1$'s. 

The $\C^2/\Z_n$ singularity can also be thought of as being produced by the collision of  $n-1$  $\C^2/\Z_2$ singularities. A $\C^2/\Z_2$ fractional collection is given by $\O_{C},\O_{C}(-1)[1]$, where $C=\P^1$. In \cite{en:fracC2} we found the following $\C^2/\Z_n$  fractional collection: $\O_{\sum_j C_j}$ and  $\O_{C_i}(-1)[1]$, with $C_i=\P^1$ from above. Moreover, we proposed a physical explanation for how the $\C^2/\Z_n$ collection is obtained from the $\C^2/\Z_2$ collections. 

It is natural  to try to extend the above procedure to an arbitrary singularity. This would also answer the more interesting question: what is the gauge theory on the world-volume of a D3-brane probing a \CY\ singularity. Phrasing it in this form, it is obvious that we are not going to get a simple answer. 

Given a singularity $S$, we can partially resolve it. This creates new singularities, call them $S_i$. Even if we started with an isolated singularity, the resolution might produce non-isolated ones (this is the case even for $\C^3/\Z_5$), therefore $i$ might be a continuous index. Viewed individually, the singularity $S_i$ has its own fractional branes $F^{S_i}_{j_i}$. As the Kahler modulus of the blow-up is turned off some of the $F^{S_i}_{j_i}$'s get destabilized, and decay. The decay products are necessarily bound states of the old $F^{S_i}_{j_i}$'s, and the ones that remain stable. We are guaranteed to find the fractional branes of the singularity $S$ among these. The question then is how do we recognize them? Is there going to be a simple physical rule like the one proposed in \cite{en:fracC2} for $\C^2/\Z_n$?

Once the fractional branes for a singularity have been obtained,  the gauge theory on the probe will be the quiver theory whose quiver is the $\Ext^1$-quiver of the fractional branes. This procedure was dubbed  ``quiver {\em LEGO}'' \cite{en:fracC2}. Using the techniques of \cite{Aspinwall:2004bs} the superpotential can also be computed, and one might obtain models of phenomenological interest. 

The organization of the paper is as follows. In Section~2 we use toric methods to investigate the geometry of the $\C^3/\Z_5$ model together with its Kahler moduli space.  Section~3 starts with a review of the Fourier-Mukai technology, and then it is applied to the various  $\C^3/\Z_5$ monodromies. In  Section~4 we use the $\Z_5$ monodromy to produce a collection of fractional branes. This collection is compared to the one given by the McKay correspondence, and we produce an interesting Seiberg duality. In  Section~5 we turn to a collection of fractional branes on the partially resolved $\C^n/\Z_m$ orbifold, using a generalization of the McKay correspondence by Kawamata. The partially resolved orbifold is singular, and it is not a global quotient either,  hence it is particularly pleasing that  we can handle it directly by geometric methods. The appendix contains some spectral sequences that are used throughout the paper.

\section{$\C^3/\Z_5$ geometries}    \label{s:c2z3}

In this section we review some aspects of the $\C^3/\Z_5$ orbifold and the associated CFT. First we work out the relevant toric geometry of $\C^3/\Z_5$, then we turn our attention to the moduli space of complexified Kahler forms, and in particular its discriminant loci. We pay particular attention to the singularities in the moduli space. 

\subsection{The toric geometry of $\C^3/\Z_5$ } 

\begin{figure}[h] 
\begin{equation}\nonumber
\begin{xy} <1.3cm,0cm>:
0*{\dir{*}}*++!D{\color{blue}v_2}="2",  (-1,-2.58)*{\dir{*}}*+!RU{\color{blue}v_1}="1"	,(1,-2.58)*{\dir{*}}*+!LU{\color{blue}v_3}="3"
,(0,-1)*{\dir{*}}="5"	,(0,-2)*{\dir{*}}="4"
\ar@{-} "1"; 0	\ar@{-} "3"; 0	\ar@{-} "1"; "3"	\ar@{-} "2";"4" 	
\ar@{-} "3"; "4"	\ar@{-}"3"; "5"	 \ar@{-} "1"; "4"	\ar@{-}"1"; "5"  	
,\POS"5"*+!L{\color{red}v_5}	,\POS"4"*++!L{\color{red}v_4}
\end{xy}
\end{equation}
  \caption{The toric fan for the resolution of the $\C^3/\Z_5$ singularity.}
  \label{f:fan}
\end{figure}
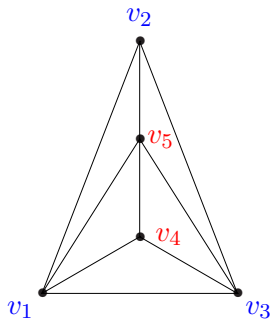

Let us consider the $\C^3/\Z_5$ variety with a supersymmetric $\Z_5$ action, i.e. $\Z_5\subset \SL(3,\Z)$. A priori there are two choice for the $\Z_5$ action:
\begin{equation}
(z_1,z_2,z_3)\mapsto (\omega z_1,\omega z_2,\omega^3 z_3)\,, \qquad \omega ^5=1\,,
\end{equation}
and 
\begin{equation}
(z_1,z_2,z_3)\mapsto (\omega^2 z_1,\omega^2 z_2,\omega z_3)\,, \qquad \omega ^5=1\,.
\end{equation}
But obviously the second one is the square of the first one, and therefore we can talk about {\em the} $\C^3/\Z_5$ variety, which is toric, and a convenient representation for it is provided by the fan in Fig.~\ref{f:fan}. 
More precisely, the fan for the $\C^3/\Z_5$ toric variety consists of only one cone, generated by the vertices $v_1$, $v_2$ and $v_3$. In this figure we also included the vertices $v_4$ and $v_5$ corresponding to the crepant resolution of the singularity. We denote the resolved space by $X$. The exceptional locus of the blow-up is reducible, and it has two irreducible components corresponding to $v_4$ and $v_5$. We denote the divisors associated to $v_i$ by $D_i$. 

The linear equivalences among the divisors and their intersections are well known, and we simply quote the result of \cite{DelaOssa:2001xk}:\footnote{Since $\C^3/\Z_5$ is non-compact, one restricts to the intersections of the compact cycles with other cycles, in this case $D_4$ and $D_5$.}
\begin{equation}\label{e:intrel}
D_4\cdot D_5=D_4\cdot D_1=h,~~ D_4\cdot D_2=0,~~D_5\cdot D_1=f, ~~
D_5\cdot D_2=h+3f
\end{equation}
and 
\begin{equation}\label{e2}
D_4^2=-3h,~~~D_5^2=-2h-5f\,.
\end{equation}
The tools of toric geometry immediately tell us that the divisor $D_4$ is a $\P^2$, while $D_5$ is the Hirzebruch surface $\F_3$. There are curves in the class of $h$ and $f$  that lie in the compact divisors $D_4$ and $D_5$. In particular $f$ is a fiber of $\F_3$, and $h$ is the $-3$ section $s$ of $\F_3$. Similarly, $h$ is the hyperplane class of $\P^2$, while $f$ does not intersect $\P^2$.

Let us spend a moment analyzing the Kahler cone, and its dual, the Mori cone. From the geometry it is clear that the curves $h$ and $f$ are the generators of the Mori cone of effective curves. Shrinking  $h$ also shrinks the divisors $D_4$, and hence gives a Type~II contraction, while shrinking  $f$ collapses the Hirzebruch surface $\F_3$ onto its base, giving a Type~III degeneration. 

The Kahler cone is dual to the Mori cone, and in our case both are two dimensional. The Kahler generators are represented by cohomology classes. We can use Poincare duality and represent the Kahler classes by 4-cycles, and in particular divisors. 
It is immediate from the intersection products in \cite{DelaOssa:2001xk} that the ordered pair $\{h,f\}$ is dual to the ordered pair $\{D_1,D_2\}$:
 \begin{equation}\label{e:ce1}
\{h,f\}\in\H_2(X,\Z)\quad \mbox{\rm is dual to}\quad \{D_1,D_2\}\in\H_4(X,\Z)\,.
\end{equation}
Therefore the Kahler cone is generated by $D_1$ and $D_2$. The precise ordering in (\ref{e:ce1}) will play an important role later on.

By Lemma 3.3.2 in \cite{Cox:Katz} there is a bijection between the Mori cone generators and the generators of the  lattice of relations of the point-set $\c A=\{v_1,\ldots, v_5 \}$. In particular, $h$ and $f$ yield the relations
\begin{equation}\label{e:chargem}
Q =
\left(\!\!
\begin{array}{rrrrr}
1&0&1&-3&1\\
0&1&0&1&-2
\end{array}
\right).
\end{equation}

\subsection{The $\C^3/\Z_5$ moduli space}   

\begin{figure}[h]
\epsfxsize=2in
\begin{center}
\epsffile{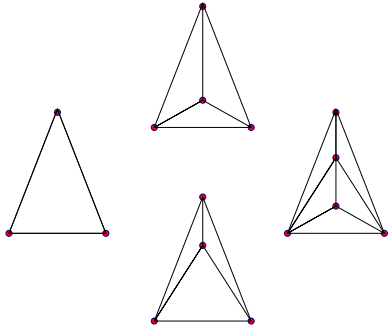}
\end{center}
  \caption{The four  triangulations of the $\C^3/\Z_5$ model.}
  \label{fig:4triang}
\end{figure}

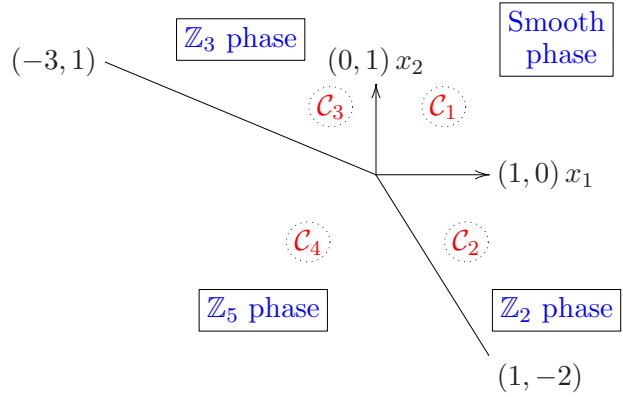
\begin{figure}[h]
\begin{equation}\nonumber
\begin{xy} <3cm,0cm>:
{\ar (0,0);(.5,0) *+!L{(1,0)\,x_1}}
,{\ar 0;(0,.6) *+!U{(0,1)\,x_2}}
,{\ar@{-} 0;(.5,-.8) *+!LU{(1,-2)}}
,{\ar@{-} 0;(-1.2,.5) *+!R{(-3,1)}}
,(.8,.6)*+[F]\txt{\color{blue} Smooth\\ \ \color{blue} phase},
,(.3,.3)*+[o][F.]{\color{red} {\mathcal C}_1}
,(-.5,-.6)*+[F]\txt{\color{blue} $\Z_5$  phase}
,(-.3,-.3)*+[o][F.]{\color{red} {\mathcal C}_4}
,(.8,-.6)*+[F]\txt{\color{blue} $\Z_2$  phase}
,(.4,-.3)*+[o][F.]{\color{red} {\mathcal C}_2}
,(-.6,.6)*+[F]\txt{\color{blue} $\Z_3$  phase}
,(-.2,.3)*+[o][F.]{\color{red} {\mathcal C}_3}
\end{xy}
\end{equation}
  \caption{The phase structure of the $\C^3/\Z_5$ model.}
  \label{fig:Z3}
\end{figure}

The point-set $\c A=\{v_1,\ldots, v_5 \}$ admits four triangulations (see Fig.~\ref{fig:4triang}). Therefore in the language of \cite{Witten:GLSM,Aspinwall:1994nu}  the gauged linear sigma model (GLSM) has four phases.\footnote{In the GLSM language the rows of this matrix represent the $\U(1)$ charges of the chiral superfields. } The secondary fan has its rays given by the columns of the matrix (\ref{e:chargem}), and is depicted in Fig.~\ref{fig:Z3}. The four phases are as follows: the completely resolved smooth phase; the two phases where one of the compact divisors $D_4$ or  $D_5$ has been blown up to partially resolve the $\Z_5$ fixed point; and finally the $\Z_5$ orbifold phase. 

The phase corresponding to the cone ${\mathcal C}_2$ can be reached from the smooth phase ${\mathcal C}_1$ by blowing down the divisor $D_5$. This creates a line of $\Z_2$ singularities in the \CY . We will refer to this phase as {\em the $\Z_2$ phase}. Similarly, the phase ${\mathcal C}_3$ is reached by blowing down the divisor $D_4$, and creates a $\C^3/\Z_3$ singularity. We call this {\em the $\Z_3$ phase}.

The orbifold points in the moduli space are themselves singular points. This fact is related to the quantum symmetry of an orbifold CFT. For the $\Z_2$ point, the homogenous coordinate ring construction of Cox \cite{Cox:HoloQout} shows a $\C^2/\Z_2$ singularity with weights $(1,-1)$. Alternatively, using the ``old'' --- group algebra $\C[\sigma^{\roof}]$ of the dual cone $\sigma^{\roof}$ ---  construction \cite{Fulton:T}, one arrives at the affine scheme $\Spec \C[y^{-1},x^2y,x] = \Spec \C[u,v,w]/(u v-w^2)$. At the $\Z_3$ point the moduli space locally is of the form $\C^2/\Z_3$, with weights $(1,2)$.

But we can take the secondary fan  in Fig.~\ref{fig:Z3} literally, since it is the fan of moduli space, viewed as a toric variety. The four cones then have natural coordinates associated to them, and these are as follows:
\begin{equation}\label{e:torcoord}
\begin{array}{|c|c|c|c|}
\hline\hline
\color{red} \cal C_1 &\color{red}  \cal C_2 &\color{red}  \cal C_3 &\color{red}  \cal C_4 \\
\hline\hline
x_1   & y_1=x_1^2x_2 & z_1=1/x_1	& w_1=1/x_1^2x_2\\
\hline
x_2  & y_2=1/x_2        & z_2= x_1x_2^3 & w_2=1/x_1x_2^3 \\
\hline
\end{array}	
\end{equation}
The coordinates $(x_1,x_2)$ are good coordinates on a dense open subset of the moduli space, but fail at the boundary divisors corresponding to the toric compactification of the moduli space. This is where one needs to make the above change of coordinates. One also notices the $x_1^2$ and $x_2^3$ terms, which reflect the $\C^2/\Z_2$ reps. $\C^2/\Z_3$ singularities. We will use these coordinates momentarily. 

\subsection{The discriminant}   

The primary components of the discriminant locus of singular CFT's can be computed using the Horn parametrization \cite{GKZ:book,MP:SumInst}. We briefly review the  construction. Let $Q=(Q_i^a)_{i=1,\ldots,n}^{a=1,\ldots,k}$ denote the matrix of charges appearing in Cox's holomorphic quotient construction \cite{Cox:HoloQout}. The primary component of the discriminant, $\Delta_0$, is a rational variety, i.e., birational to a projective space.
Horn uniformization gives an explicit rational parametrization for $\Delta_0$. Accordingly, we introduce $k$ auxiliary variables, $s_1,\ldots,s_k$, and form the linear combinations
\begin{equation}\label{e:hu1}
\xi_i=\sum_{a=1}^k Q_i^a\,s_a\,,\qquad \mbox{for all $i=1,\ldots,n$}\,.
\end{equation}
Let $(x_a)_{a=1,\ldots,k}$ be local coordinates on the moduli space of complexified Kahler forms, or equivalently, on the complex structure moduli space of  the mirror. $\Delta_0$ then has the following parameterization:
\begin{equation}\label{e:hu2}
x_a=\prod_{i=1}^n \xi_i^{\;Q_i^a}\,,\qquad \mbox{for all $a=1,\ldots,k$}\,.
\end{equation}

In our context the matrix of charges in question is $Q$ from Eq.~(\ref{e:chargem}), and $(x_1,x_2)$ are the local coordinates on the moduli space corresponding to the large radius phase. Applying the Horn uniformization equations, (\ref{e:hu1}) and (\ref{e:hu2}), gives 
\begin{equation}\label{e:cpxc}
x_1=-\frac{s_1^2(s_1-2s_2)}{(3s_1-s_2)^3}\,, \qquad x_2=-\frac{s_2(3s_1-s_2)}{(s_1-2s_2)^2}\,.
\end{equation}
Both equations are homogeneous, and therefore $x_1$ and $x_2$ depend only on the ratio $s_1/s_2$. Eliminating $s_1/s_2$ gives the sought after equation for $\Delta_0$:
\begin{equation}\label{eq:dZ3}
\Delta_0 = 3125 {{x}}_{1}^{2} {{x}}_2^{3}+500 {{x}}_{1} {{x}}_2^{2}-225 {{x}}_{1} {{x}}_2+16 {{x}}_2^{2}+27 {{x}}_{1}-8 {{x}}_2+1	\,.
\end{equation}
In fact this is the only component of the primary discriminant.

The discriminant curve itself is singular. It has two singular points: 
\begin{equation}
(x_1,x_2)=\left(-\ff1{25}, \ff15	\right)\quad \mbox{and }\quad (x_1,x_2)=\left(-\ff{32}{675}, \ff9{20}	\right)\,.
\end{equation}
The first singular point is a double root of the $\Delta_0 =grad\, \Delta_0 =0$ equation, while the second is a triple root. This hints to the fact that the local structure of the two singular points is different.

To see the nature of the singularity we need to choose convenient coordinates around the singularities. We treat the point $\left(-\ff{32}{675}, \ff9{20}\right)$ first. A convenient change of variables is the following:
\begin{equation}
(x_1,x_2)=\left(-\ff{32}{675}+\ff8{135}y_1,\ff9{20}+\ff34 y_2\right)\,.
\end{equation}
In terms of the new variables the discriminant becomes
\begin{equation}
\begin{split}
\Delta_0 = &{y_1}^{2}+{y_2}^{2}+2y_1y_2+\ff{10}3y_1{y_2}^{
2}-{\frac {200}{27}}y_1{y_2}^{3}\\
&+5{y_1}^{2}y_2+{
\frac {25}{3}}{y_1}^{2}{y_2}^{2}+{\frac {125}{27}}{y_1
}^{2}{y_2}^{3}+{\frac {80}{27}}{y_2}^{3}\,.
\end{split}
\end{equation}
This suggests a further change of variables: $(y_1,y_2)\mapsto (y_1-y_2,y_2)$, in terms of which the leading terms of $\Delta_0$ are
\begin{equation}
\Delta_0 = {y_1}^{2}-{\frac {20}{3}}\,y_1{y_2}^{2}+{\frac {125}{27}}
\,{y_2}^{3}+5{y_1}^{2}y_2
\end{equation}
which shows that $\Delta_0$ has a cusp at $(x_1,x_2)=\left(-\ff{32}{675}, \ff9{20}\right)$.

A similar analysis can be performed at the other singularity, $(x_1,x_2)=\left(-\ff1{25}, \ff15	\right)$, and show that $\Delta_0$ has an ordinary double point there. The fact that one of the singularities is a cusp will be important later on, in Section \ref{s:loops}, as it allows for different monodromies around different parts of the discriminant.

\subsection{Intersections}    \label{s:intm}

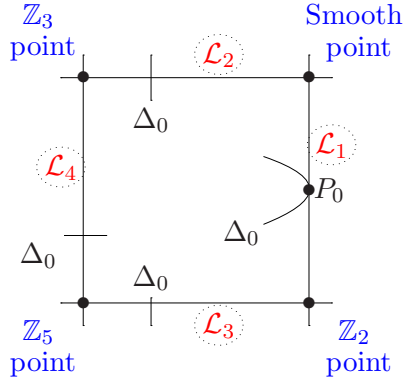
\begin{figure}[h]
\begin{equation}\nonumber
\begin{xy} <3cm,0cm>:
{\ar@{-} (-.1,0);(1.1,0) }	,{\ar@{-} (1,.1);(1,-1.1) }	,{\ar@{-} (1.1,-1);(-.1,-1)}	,{\ar@{-} (0,.1);(0,-1.1)} 
,(1.2,.2)*\txt{\color{blue} Smooth\\ \ \color{blue} point}
,(.6,.1)*+[o][F.]{\color{red} {\mathcal L}_2}
,(1.2,-1.2)*\txt{\color{blue} $\Z_2$\\ \ \color{blue} point}
,(1.1,-.3)*+[o][F.]{\color{red} {\mathcal L}_1}
,(-.2,-1.2)*\txt{\color{blue} $\Z_5$\\ \ \color{blue} point}
,(.6,-1.1)*+[o][F.]{\color{red} {\mathcal L}_3}
,(-.2,.2)*\txt{\color{blue} $\Z_3$\\ \ \color{blue} point}
,(-.1,-.4)*+[o][F.]{\color{red} {\mathcal L}_4}
,{\ar@{-} (0.3,0.1);(0.3,-.1)   *+!U{\Delta_0}}		,{\ar@{-} (0.3,-1.1);(0.3,-.8)   *+!U{\Delta_0}}			
,{\ar@{-} (.1,-.7);(-.2,-.7)   *+!U{\Delta_0}}		,(0.7,-.6)   *+!U{\Delta_0}
,{(.8,-.35);(.8,-.65) **\crv{(1.091,-.45) & (1.05,-.55)}}
,(0,0)*{\bullet}, (0,-1)*{\bullet} 	,(1,0)*{\bullet} ,(1,-1)*{\bullet} 
,(1.09,-.5)*{P_0}	,(1,-.5)*{\bullet}	
\end{xy}
\end{equation}
  \caption{The moduli space of the $\C^3/\Z_5$ model.}
  \label{fig:modsp}
\end{figure}

We now turn to a question which will be of crucial importance. We have already seen that the four maximal cones ${\mathcal C}_1$, ... , ${\mathcal C}_4$ in Fig.~\ref{fig:Z3} correspond to the four distinguished phase points. Similarly, the four rays correspond to curves in the moduli space. It is immediate to see these are all $\P^1$'s, at least topologically. For us the analytic structure is important, and we need to be more careful here. The four curves connecting the different phase points are sketched in Fig.~\ref{fig:modsp}. The curves in question are weighted projective lines: ${\mathcal L}_1=\P^1(1,2)$ ${\mathcal L}_2=\P^1(1,3)$,  ${\mathcal L}_3=\P^1(2,5)$  and ${\mathcal L}_4=\P^1(3,5)$. Since the singularities are in codimension one, these spaces in fact are not singular, and they are all isomorphic to $\P^1$.

In terms of the coordinates $(x_1,x_2)$ we have ${\mathcal L}_1\!: (x_1=0)$ and ${\mathcal L}_2\!: (x_2=0)$. The discriminant $\Delta_0$ intersects these two lines. To see  what the order of intersection is  observe that:
\begin{equation}
\Delta_0 (x_1=0)= (4x_2-1)^2\,,\qquad
\Delta_0 (x_2=0)= 27x_1+1\,.
\end{equation}
Therefore $\Delta_0$ intersects ${\mathcal L}_1$ tangentially, while it is transverse to  ${\mathcal L}_2$. We depicted this fact in Fig.~\ref{fig:modsp} using a parabola and resp. a short segment.

But what about the intersection of $\Delta_0$ with ${\mathcal L}_3$  and ${\mathcal L}_4$? Clearly $(x_1,x_2)$ are not the proper coordinates in these patches, but we can use the correct ones from Eq.~(\ref{e:torcoord}). Let's focus on ${\mathcal L}_3$ first. In the coordinates $(y_1,y_2)$ of the cone ${\mathcal C}_2$, ${\mathcal L}_3$ is given by the equation $y_2=0$. To determine the form of $\Delta_0$ in the new coordinates is more complicated, but there two different ways of doing it. The naive ways is to observe that $y_1y_2=x_1^2$, while $y_2=1/x_2$. So one can solve for $x_1$ and $x_2$, in terms of $y_1$ and $y_2$, and substitute into $\Delta_0$, and then try to square an appropriately rearranged expression to get rid of the $\sqrt{y_1y_2}$ terms.

The more efficient method is to go back to the Horn parametrization, and run it for the coordinates $(y_1,y_2)$. This amounts to replacing the matrix (\ref{e:chargem}) with 
\begin{equation}\label{e:chargem1}
Q =
\left(\!\!
\begin{array}{rrrrr}
2 &1&  2 &-5 &0 \\
0 &-1& 0 &-1& 2 
\end{array}
\right)\,.
\end{equation}
Both methods give the same answer:
\begin{equation}
\begin{split}
\Delta_0 =& (y_2-4)^4+5^{10}y_1^{2}\\
& \left( 10^5-3\cdot 10^5 y_2+2\cdot 5^5 \cdot37{y_2}^{2}-23\cdot 15^3{y_2}^
{3}+2\cdot 3^5 \cdot 5^{2} {y_2}^{4}-3^6{y_2}^{5} \right) y_1\,.
\end{split}
\end{equation}
One immediately has that
\begin{equation}
\Delta_0 (y_1=0)= (y_2-4)^4\,,\qquad
\Delta_0 (y_2=0)= (3125 y_1 + 16)^2\,.
\end{equation}
This would suggest a fourth order and a tangential intersection, but in fact we have to remember that the $(y_1,y_2)$ coordinates double all the intersections, and hence we have a tangential and a transverse intersection. 

To shed more light on the last statement, let us recall that the curve ${\mathcal L}_1$,  given by the equation $x_1=0$, after the change of coordinates $ y_1=x_1^2x_2$, becomes
$y_1=0^2$, exhibiting the doubling that we referred to before. The tangential intersection at $y_2=4$ is the previous $x_2=1/4$ point, which was also a  tangential intersection. The curve $y_2=0$ is none other than ${\mathcal L}_3$, and hence it is transverse to $\Delta_0$.

Finally we turn to the   intersection of ${\mathcal L}_4$ and $\Delta_0$.  In this case the naive elimination method fails due to the cube-root, but the Horn parametrization works. A similar analysis shows that the curve ${\mathcal L}_4$ given by  the equation $z_1=0$ is also  transverse to $\Delta_0$. 

\section{$\C^3/\Z_5$ monodromies}    \label{s:m}

We start this section with a brief review of Fourier-Mukai functors. Then we  express the various  monodromy actions on D-branes in terms of Fourier-Mukai equivalences. The remaining part of the section deals with expressing the D-brane monodromies appearing in the $\C^3/\Z_5$ moduli space in terms of Fourier-Mukai functors.

\subsection{Fourier-Mukai functors}\label{s:fmf1}

For the convenience of the reader we review some of the key notions concerning Fourier-Mukai functors, and at same time specify the conventions used. We will make extensive use of this technology in the rest of the paper.  Our notation follows \cite{Horj:EZ}.  

Given two non-singular proper algebraic varieties, $X_1$ and $X_2$, an object ${\cal K} \in \D(X_1\! \times\! X_2)$ determines a functor of triangulated categories $\Phi_{\cal K}\!: \D(X_1) \to \D(X_2)$ by the formula\footnote{
$\D(X)$ denotes the {\em bounded} derived category of coherent sheaves on $X$. $\R p_{2*}$ is the total right derived functor of $p_{2*}$, i.e., it is an exact functor from $\D(X)$ to $\D(X)$. Similarly, $\Ltensor$ is the total left derived functor of $\otimes$. In later sections these decorations will be subsumed.}
\begin{equation}
\Phi_{\cal K}(A):=\R p_{2*} \big(\,{\cal K} \Ltensor p_1^*(A)\,\big)\,,
\end{equation}
where $p_i\!: X\! \times\! X \to X$ is projection to the $i$th factor:
\begin{equation}
\xymatrix{
  &X_1 \!\times\! X_2\ar[dl]_{p_1}\ar[dr]^{p_2}&\\
  X_1 & & X_2\,.}
\end{equation}
The object ${\cal K} \in \D(X_1\!\times\! X_2)$ is called the {\bf kernel} of the Fourier-Mukai functor $\Phi_{\cal K}$.

It is convenient to introduce the {\bf external tensor product} of two objects $A\in\D(X_1)$ and $B\in\D(X_2)$ by the formula
\begin{equation}
A\boxtimes B=p_2^*A\Ltensor p_1^*B\,.
\end{equation}

The importance of Fourier-Mukai functors when dealing with derived categories stems from the following theorem of Orlov:\footnote{Theorem 2.18 in \cite{Orlov:96}. The theorem has been generalized for smooth quotient stacks associated to normal projective varieties \cite{Kawamata:DC}.}
\begin{thm} \label{thm:orlov}
Let $X_1$ and $X_2$ be smooth projective varieties.
Suppose that $\mathsf{F}\!: \D(X_1)\to\D(X_2)$ is an equivalence of triangulated categories. Then there exists an object $\c K\in \D(X_1\!\times \! X_2)$, unique up to isomorphism, such that the functors $\mathsf{F}$ and $\Phi_{\c K}$ are isomorphic.
\end{thm}

The first question to ask is how to compose Fourier-Mukai (FM) functors. Accordingly, let $X_1$ $X_2$ and $X_3$ be three non-singular varieties, while let ${\cal F} \in \D(X_1\! \times\! X_2)$ and ${\cal G} \in \D(X_2\! \times\! X_3 )$ be two kernels. Let $p_{i j}\colon X_1\! \times\! X_2\! \times\! X_3\to X_i\! \times\! X_j$ be the projection map. A well-known fact is the following:
\begin{prop}\label{prop1}
The composition of the functors $\Phi_{\cal F}$ and $\Phi_{\cal G}$  is given by the formula
\begin{equation}
\Phi_{\cal G}\comp \Phi_{\cal F} \simeq\Phi_{\cal H}\,,\quad {\rm where}\quad
{\cal H}=\R p_{13*} \big(\, p_{23}^* ( {\cal G})\Ltensor  p_{12}^* ({\cal F})\big)\,.
\end{equation}
\end{prop}

Prop.~\ref{prop1} shows that  composing two FM functors gives another FM functor, with a simple kernel.
The composition of the kernels ${\cal  F}$ and ${\cal G} \in \D(X\times X)$ is therefore defined as
\begin{equation} \label{def:comp}
{\cal G}\star {\cal F} := \R p_{13*}\big(\, p_{23}^* ({\cal G})\Ltensor  p_{12}^* ({\cal F})\, \big).
\end{equation}

There is an identity element for the composition of kernels:
$\delta_* (\O_X),$ where $\delta : X \hookrightarrow X\! \times\! X$
is the diagonal embedding. For brevity we will denote $\delta_* (\O_X)$ by $\O_\Delta$:
\begin{equation}
\O_\Delta:=\delta_* (\O_X)\,.
\end{equation}
$\O_\Delta=\delta_* (\O_X)$ has the expected properties:
\begin{equation}\label{fm:id}
\O_\Delta\star{\cal G }={\cal G }\,\star\O_\Delta={\cal G }\,, \quad \mbox{for all ${\cal G} \in \D(X\times X)$}.
\end{equation}

Finally, the functors
\begin{equation}\label{trig:mor}
\begin{split}
&\Phi_{23}\!: \D(X_1\! \times\! X_2)\to \D(X_1\! \times\! X_3),\quad
 {\cal G }_{23}\in\D(X_2\! \times\! X_3),\quad
  \Phi_{23} (-):= {\cal G }_{23} \star -\,, \\
&\Phi_{12}\!: \D(X_2\! \times\! X_3)\to \D(X_1\! \times\! X_3),\quad
 {\cal G }_{12}\in\D(X_1\! \times\! X_2),\quad
  \Phi_{12} (-):= - \star {\cal G }_{12}\,,
\end{split}
\end{equation}
are morphisms between triangulated categories, i.e., they {\em preserve} distinguished triangles.

The composition of kernels is also associative
\begin{equation}\label{eq:assoc}
{\cal G}_{3} \star ({\cal G}_{2} \star
{\cal G}_{1}) \cong ({\cal G}_{3} \star {\cal G}_{2}) \star {\cal G}_{1}\,.
\end{equation}

Now we have all the technical tools ready to study the monodromy actions of physical interest.

\subsection{Monodromies in general}

The moduli space of CFT's contains the moduli space of Ricci-flat Kahler metrics. This, in turn, at least locally has a product structure, with the moduli space of Kahler forms being one of the factors. This is the moduli space of interest to us. In what follows we study the physics of D-branes as we move in the moduli space of complexified Kahler forms. This space is a priori non-compact, and its compactification consists of two different types of boundary divisors. First we have the {\em large volume} divisors.  These correspond to certain cycles being given infinite volume. The second type of boundary divisors are the irreducible components of the {\em discriminant}. In this case the CFT becomes singular. Generically this happens because some D-brane (or several of them, even infinitely many) goes massless at that point, and therefore the effective CFT description brakes down. For the quintic this is the well known conifold point.

The monodromy actions around the above divisors are well understood. We will need a more abstract version of this story, in terms of Fourier-Mukai functors, which we now recall.\footnote{For an extensive treatment of monodromies in terms of Fourier-Mukai functors see \cite{en:Horja}.} 

Large volume monodromies are shifts in the $B$ field: ``$B\mapsto B+1$''. If the Kahler cone is higher dimensional, then we need to be more precise, and specify a two-form, or equivalently a divisor $D$. Then the monodromy becomes $B\mapsto B+D$. We will have more to say about the specific $D$'s soon. 

The simplest physical effect of this monodromy on a D-brane is to shift its charge, and this translates in the Chan-Paton language into tensoring with the line bundle $\O_X(D)$. This observation readily extends to the derived category:
 \begin{prop}\label{p:lr} 
The large radius monodromy associated to the divisor $D$ is
\begin{equation}
   \ms{L}_{D}(\mathsf{B}) = \mathsf{B}\Ltensor \O_X(D)\,,\qquad \mbox{for all $\mathsf{B} \in \D(X)$}\,.
\end{equation}
Furthermore, this is a Fourier-Mukai functor $\Phi_{{\cal L}}$, with kernel 
\begin{equation}
{\cal L}=\delta_*\O_X(D)\,,
\end{equation}
where  $\delta \!: X \hookrightarrow X\! \times\! X$ is again the diagonal embedding.\footnote{For a proof of this statement we refer to \cite{en:fracC2}.}
\end{prop}

For the conifold-type monodromies one has the following conjecture:\footnote{
Originally due to Kontsevich, Morrison  and Horja, and presented in \cite{en:Horja}.}
\begin{conj}\label{conj:a}
If we loop around a component of the discriminant locus associated with a single D-brane $\mathsf{A}$ becoming
massless, then this results in a relabeling of D-branes given by the autoequivalence of the derived category $\D(X)$
\begin{equation}
\mathsf{B}\longmapsto
\Cone{\RHom_{\D(X)}(\mathsf{A},\mathsf{B})\Ltensor\mathsf{A}\longrightarrow\mathsf{B}}.
\end{equation}
\end{conj}

This action is again  of Fourier-Mukai type. Lemma 3.2 of \cite{ST:braid} provides us with the following simple relation for any $\mathsf{B}\in\D(X)$:
\begin{equation}
\Phi_{ \Cone{ \mathsf{A}^{\roof}\boxtimes\,\mathsf{A} \to \O_\Delta} }(\mathsf{B}) \iso
\Cone{ \R\!\Hom_{\D(X)}(\mathsf{A},\mathsf{B})\Ltensor\mathsf{A}\to\mathsf{B} },
\end{equation}
where for an object $\mathsf{A}\in\D(X)$ its dual is defined by
\begin{equation}
\mathsf{A}^{\roof}= \R\!\!\sHom_{\D(X)}( \mathsf{A}, \O_X).
\end{equation}
Throughout the paper $\Cone{f}$ refers to the cone of the morphisms $f\colon A\to B$.

Since the functor $\Phi_{ \Cone{ \mathsf{A}^{\roof}\boxtimes\,\mathsf{A}\to\O_\Delta } }$ will play a crucial role, we give it a name:
\begin{equation}\label{e:refl}
\ms{T}_{\mathsf{A}} :=
\Phi_{ \Cone{\mathsf{A}^{\roof}\boxtimes\,\mathsf{A}\to\O_\Delta } }\,,\quad
\ms{T}_{\mathsf{A}}(\mathsf{B}) =\Cone{ \R\!\Hom_{\D(X)}(\mathsf{A},\mathsf{B})\Ltensor\mathsf{A}\to\mathsf{B} }.
\end{equation}

The question of when is $\ms{T}_{\mathsf{A}}$ an autoequivalence has a simple answer. For this we need the following definition:
\begin{definition} \label{def:spherical}
Let $X$ be smooth projective \CY\ variety of dimension $n$. An object $\mathsf{E}\in \D(X)$ is called {\em n-spherical} if $\Ext^r_{\D(X)}(\mathsf{E},\, \mathsf{E})$ is equal to $\H^r(S^n,\,\C)$, that is $\C$ for $r = 0,n$ and zero otherwise.
\end{definition}

One of the main results of \cite{ST:braid} is the following theorem (Prop. 2.10):
\begin{thm}
If $\mathsf{E}\in \D(X)$ is n-spherical then the functor $\ms{T}_{\mathsf{E}}$ is an autoequivalence.
\end{thm}

This brief review brings us to a point where we can apply this abstract machinery to study the $\C^3/\Z_5$ monodromies, and eventually use them to construct the fractional branes.

\subsection{$\C^3/\Z_5$ monodromies}	\label{s:loops}

Now we have all the ingredients necessary for constructing the monodromy actions needed to generate the fractional branes. The toric fan for the moduli space of complexified Kahler forms was depicted in Fig.~\ref{fig:Z3}. The four maximal cones ${\mathcal C}_1$, ... , ${\mathcal C}_4$ correspond to the four distinguished phase points. The four edges correspond to curves in the moduli space. As discussed in Section~\ref{s:intm} the curves in question are weighted projective lines: ${\mathcal L}_1=\P^1(1,2)$,  ${\mathcal L}_2=\P^1(1,3)$,  ${\mathcal L}_3=\P^1(2,5)$  and ${\mathcal L}_4=\P^1(3,5)$, and are in  fact all isomorphic to $\P^1$.  The discriminant $\Delta_0$ intersects the four lines, and we analyzed the order of every intersection. All this is summarized in Fig.~\ref{fig:modsp}. 

When talking about monodromy there are two cases to be considered. One can loop around a divisor, i.e., real codimension two objects; or one can loop around a point inside a complex curve. Of course the two notions are {\em not} unrelated. Our interest will be in the second type of monodromy: looping around a point inside a $\P^1$.

What we would like to write down is the monodromy inside ${\mathcal L}_3$ or ${\mathcal L}_4$ around the $\Z_5$ point. Since there is no direct approach to doing this, we follow an indirect way: ``go around''. Both ${\mathcal L}_1$ and ${\mathcal L}_3$ are spheres, with three marked points, and we can compute the corresponding monodromies. Our approach is to go from the smooth point to the $\Z_5$ point by first ``moving'' inside  ${\mathcal L}_1$ and then ${\mathcal L}_3$. An equally valid path is through ${\mathcal L}_2$ and ${\mathcal L}_4$, but we are not going to deal with this.

We start with ${\mathcal L}_1$, which has three distinguished points: the smooth point, $P_0={\mathcal L}_1\cap \Delta_0$ and the $\Z_2$ point. Monodromy around the smooth point inside ${\mathcal L}_1$ is a large radius monodromy, and (\ref{e:ce1}) together with Prop.~\ref{p:lr} tell us that it is precisely $\ms{L}_{D_2}$. 

What about the monodromy around $P_0={\mathcal L}_1\cap \Delta_0$? $P_0$ is a  conifold-type point, but the intersection is tangential at this point. We know that at this point every D5-brane wrapping the fibers of the shrinking cycle $D_5$ should go massless. The mass depends on the central charge, which in turn is only a function of the K-theory class. Using the K-theory analysis of \cite{DelaOssa:2001xk} one can indeed verify the masslessness of the fiber-wrapping D5-branes.
 
In other words, there are infinitely many branes going massless at this point, and Conjecture~\ref{conj:a} does {\em not} apply! At this point we could use Conjecture~4 from \cite{en:Horja}, developed for general degenerations, but we find it easier to proceed by an indirect method, which borrows from the techniques that gave supporting evidence for  Conjecture~4 in \cite{en:Horja}.

Looking at Fig.~\ref{fig:modsp}, it is clear that there should be a relation between the following three monodromies: 
monodromy around $P_0$ inside ${\mathcal L}_1$; monodromy around $\Delta_0$; and finally, monodromy around ${\mathcal L}_1$.

We can determine this relationship using knot theory. We remind the reader some facts about the topology of plane curve singularities. Let the curve $C$ be given by the vanishing locus of the irreducible polynomial $f$. We assume that the origin is an isolated singularity of $C$. This information is distilled by the notation: ``$(C,0)\!:\, f=0$''. It is customary to take a small 3-sphere $S^3_{\!\epsilon}$ of radius ${\epsilon}$ around the origin and look at the intersection ${\bf L}(C,0):=C \cap S^3_{\!\epsilon}$. $L(C,0)$ is in fact a knot, and is called the {\em link of the singularity}.

For $K_1$ and $K_2$ two disjoint knots in $S^3$ their {\em linking number } $\sf{lk}(K_1, K_2)$ is defined as follows: choose a 2-chain $S$ in $S^3$ such that $\partial S=K_1$ (this is possible since $\H_1(S^3)=0$). $\sf{lk}(K_1, K_2)$ is then given by the intersection number $S\cdot K_2$.\footnote{We are being intentionally loose about the orientation, since this won't affect our subsequent computations.
}

It is probably not surprising that the linking number of two algebraic knots can be computed purely algebraically. For convenience we reproduce Proposition II.2.12 of \cite{Dimca}:
\begin{prop}
Let $(C_i,P_0)\!:\, f_i=0$ be two distinct irreducible plane curve singularities. The linking number of their knots equals their order of intersection:
\begin{equation}
\sf{lk}\big({\bf L}(C_1,P_0), {\bf L}(C_2,P_0)\big)=C_1\cdot_{P_0} C_2\,,
\end{equation}
where $C_1\cdot_0 C_2$ is the local intersection number of $C_1$ and $C_2$.
\end{prop}

\begin{figure}[h]
\begin{equation}\nonumber
\knotholesize{3mm}
\begin{xy} <10mm,0mm>: 
  ,{\hcap[-4] \vover \vcross \vcross \vover-}
  , +(1,4), {\hcap[4]}
,(-1,-.3)*+[F]{\color{red}{{\mathcal L}_1 \cap S^3_{\!\epsilon}}}	
,(4 ,-.3)*+[F]{\color{red}{ \Delta_0 \cap S^3_{\!\epsilon}}}
,{(-1.5,-2)*{}="a" ,\ar@/_4pt/@*{[|(2)]}_(0.7){\color{blue}\mbox{\Large b}} "a"+(1,0); "a"}
,{(3.5,-2 )*{}="a"  ,\ar@/_4pt/@*{[|(2)]}_(0.3){\color{blue}\mbox{\Large a}} "a"+(1,0); "a"}
,{(.6,-1)*     {}="a" ,\ar@/^6pt/@*{[|(2)]}^(0.8){\color{blue}\mbox{\Large $a_2$}} "a"; "a"+(1,0)}
,{(1.5,-3.2)*{}="a" ,\ar@/^7pt/@*{[|(2)]}^(0.8){\color{blue}\mbox{\Large $a_1$}} "a"; "a"+(1,0)}
\end{xy}  
\end{equation}
  \caption{The two links: ${\bf L}({\mathcal L}_1,P_0)={\mathcal L}_1 \cap S^3_{\!\epsilon}$ and ${\bf L} (\Delta_0,P_0)=\Delta_0 \cap S^3_{\!\epsilon}$.}
  \label{f:}
\end{figure}
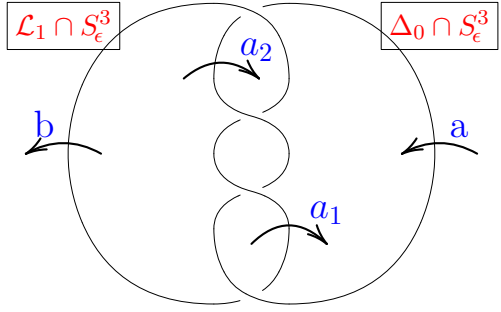

We can apply readily the proposition to our case, for $\Delta_0$ and ${\mathcal L}_1$. For us $P_0={\mathcal L}_1\cap \Delta_0$. We already know that $\Delta_0\cdot_{P_0} {\mathcal L}_1=2$.Both ${\bf L}({\mathcal L}_1,P_0)= {\mathcal L}_1 \cap S^3_{\!\epsilon}$ and ${\bf L}(\Delta_0,P_0)=\Delta_0 \cap S^3_{\!\epsilon}$ are unknotted circles;\footnote{For the general result see for example Proposition 4.2.21 in \cite{Dimca}.
}
and the link is represented in Figure~\ref{f:}.

We can use the Wirtinger representation to compute a presentation for the fundamental group of the complement of the link.\footnote{For a nice reference see \cite{Rolfsen}.
}
Fix a basepoint above the sheet of paper, just before the eyes of the reader. Each arrow in Fig.~\ref{f:} represents a loop with that basepoint going {\em under} the given arc; and therefore representing an element of $\pi_1$. The four crossing relations from top to bottom are:\footnote{Our order convention is that $x\,y$ represents the path $x$
followed by the path $y$.
}
\begin{equation}
b\,a=a_2\, b=a_1\,a_2=a\,a_1=b\,a\,.
\end{equation}
This is three independent relations among the four loops $b,\,a,\,a_1,\,a_2$. We can solve for $a_1$ and $a_2$: $a_1=a^{-1}b\, a$ and $a_2=b\, a\, b^{-1}$, and also get one relation
\begin{equation}\label{e:relkn}
b\,a\,b\,a=a\,b\,a\,b\,.
\end{equation}
Thus the fundamental group of the complement of the two knots is the group on two generators $(a,b)$ subject to the single relation (\ref{e:relkn}):
\begin{equation}
\pi_1\big(S^3-(L_1\cup L_2)\big)=\langle a ,b\;|\; b\,a\,b\,a=a\,b\,a\,b \rangle\,.
\end{equation}

Now remember that we are after the loop that encircles $P_0$ inside ${\mathcal L}_1$. But this is the counter-clockwise path ${\mathcal L}_1 \cap S^3_{\!\epsilon}$ from Fig.~\ref{f:}, and is homotopic to $a\,a_2= a\,b\, a\,b^{-1}$. Using (\ref{e:relkn}) this also equals $ b^{-1}\,a\,b\, a$. But $a$ is a loop around $\Delta_0$ and by Conjecture~\ref{conj:a} the associated monodromy is $\ms{T}_{j_*\O_{D_5}}$. Similarly $b$ is a loop around ${\mathcal L}_1$, and  Eq.~(\ref{e:chargem}) tells us that it is $\ms{L}_{D_1}$. It thus follows that the monodromy around $P_0$ inside ${\mathcal L}_1$ is
\begin{equation} \label{eq:pf}
\ms{M}_{P_0}\, = \,
\ms{L}_{D_1}^{-1}\comp \,  \ms{T}_{j_*\O_{D_5}}\comp \, \ms{L}_{D_1} \comp \, \ms{T}_{j_*\O_{D_5}}\,.
\end{equation}
Putting the pieces together we have the monodromy around the $\Z_2$ point inside ${\mathcal L}_1$:
\begin{equation}
\ms{M}_{\Z_2}\, = \,
\ms{L}_{D_1}^{-1}\comp \,  \ms{T}_{j_*\O_{D_5}}\comp \, \ms{L}_{D_1} \comp \, \ms{T}_{j_*\O_{D_5}}  \comp \, \ms{L}_{D_2}\,.
\end{equation}
We can simplify this expression by observing that 
$$\ms{L}_{D_1}^{-1}\comp \,  \ms{T}_{j_*\O_{D_5}}\comp \, \ms{L}_{D_1}\iso\ms{T}_{j_*\O_{D_5}(-f)},$$
and have
\begin{equation}\label{e:m1} 
\ms{M}_{\Z_2}\, = \,
\ms{T}_{j_*\O_{D_5}(-f)} \, \comp \, \ms{T}_{j_*\O_{D_5}}  \comp \, \ms{L}_{D_2}\,.
\end{equation}

A much simpler computation shows that the monodromy around the $\Z_3$ point inside ${\mathcal L}_2$ is
\begin{equation}\label{e:m1z} 
\ms{M}_{\Z_3}\, = \, \ms{T}_{i_*\O_{D_4}} \comp \, \ms{L}_{D_1}  \,.
\end{equation}

Now we can  continue our march towards the $\Z_5$ point inside ${\mathcal L}_3$. Once again there are three distinguished points: the $\Z_5$ point, ${\mathcal L}_3\cap \Delta_0$ and the $\Z_2$ point. By the same token as before, monodromy around ${\mathcal L}_3\cap \Delta_0$ is $\ms{T}_{i_*\O_{D_4}}$. Of course we want the  monodromy around the $\Z_5$ point, so we need the monodromy around this $\Z_2$ point, which a priori has nothing to do with the previous $\Z_2$ monodromy inside ${\mathcal L}_1$. But the monodromy around the $\Z_2$ point is more subtle. Fig.~\ref{fig:modsp} in fact is quite misleading, since in reality the spheres ${\mathcal L}_1$ and ${\mathcal L}_3$ intersect transversely in 4-space. Moreover, the intersection point is an orbifold itself: $\C^2/\Z_2$. To see what happens, we need to work out the local fundamental group of the complement of ${\mathcal L}_1$ and ${\mathcal L}_3$.\footnote{This situation is similar to the one analyzed in \cite{navigation}.}

Since the intersection point is a $\C^2/\Z_2$ orbifold, we surround it by the {\em lens space} $L=S^3/\Z_2$, instead of the usual sphere $S^3$.  ${\mathcal L}_1$ and ${\mathcal L}_3$ are both smooth curves and therefore intersect $L$ in unknotted circles. This way we reduced the problem of computing $\pi_1(\C^2/\Z_2-\{{\mathcal L}_1\cup {\mathcal L}_3\})$ to computing $\Pi=\pi_1(L-\{{\mathcal L}_1\cup {\mathcal L}_3\})$. To evaluate this consider the covering map $q\!:S^3\to L$, with free $\Z_2$ action, induced by the $\Z_2$ action on $\C^2$. The intersection of both ${\mathcal L}_1$ and ${\mathcal L}_3$ with $L$ lift under $q^{-1}$ to unknotted circles in $S^3$. These circles are linked once and thus $\pi_1(S^3-\{q^{-1}({\mathcal L}_1)\cup q^{-1}({\mathcal L}_3)\})=\Z\oplus\Z$.\footnote{An equivalent way of seeing this is to note that $\C^2-\{{\mathcal L}_1\cup {\mathcal L}_3\}$ is homotopic to $\C^*\times \C^*$, and that $S^1\times S^1$ is a deformation retract of $\C^*\times \C^*$. } 
The generators are the loops around $q^{-1}({\mathcal L}_1)$ and $q^{-1}({\mathcal L}_3)$, we call them $g_1$ and $g_2$.

Since $q$ is a normal cover we have a short exact sequence of abelian groups
\begin{equation} \label{eq:ext3}
\ses{\Z\oplus\Z}{\Pi}{\Z_2}\,.
\end{equation}
We can easily show that $\Pi=\Z\oplus\Z$ as well, by choosing a convenient fundamental domain, and two generators for $\Pi$: $l_1$ encircles ${\mathcal L}_1$, while $l_2$ goes from a basepoint to its antipodal. This second generator is a closed curve in  $L=S^3/\Z_2$ because of the quotienting, but it does not lift to $S^3$. Nevertheless, $2l_2$ {\em does} lift to $S^3$, and $q^{-1}(2l_2)=g_1+g_2$. In term of the two basis $\left\langle g_1,g_2\right\rangle $ and $\langle l_1,l_2\rangle $ we have the non-trivial map in (\ref{eq:ext3}):
\begin{equation}
\xymatrix@1{\Z\oplus\Z 
\ar[rrr]^-{\left(\!\!
\begin{array}{rr}
1&-1\\
0&2
\end{array}
\right)} 
&&& \Z\oplus\Z}\,.
\end{equation}

Now we can continue our monodromy calculation. We claim that the loop around ${\mathcal L}_1\cap {\mathcal L}_3$ inside ${\mathcal L}_1$ is {\em homotopic} to the loop around ${\mathcal L}_1\cap {\mathcal L}_3$ inside ${\mathcal L}_3$. This statement is not to be taken literally though. Neither ${\mathcal L}_1$ nor ${\mathcal L}_3$ are part of the moduli space, so we are not looping inside them. What we have are loops that are infinitesimally close to such loops, but  lie outside ${\mathcal L}_1$ or ${\mathcal L}_3$. This distinction is usually irrelevant, but for us the singularity brings it to the forefront. What we need to do is to deform the loop inside ${\mathcal L}_1$ around ${\mathcal L}_1\cap {\mathcal L}_3$ so that it doesn't intersect ${\mathcal L}_1$ or ${\mathcal L}_3$, and similarly for the loop inside ${\mathcal L}_3$ around ${\mathcal L}_1\cap {\mathcal L}_3$. The reader can convince himself that the generic deformations are indeed both  homotopic to $l_2$. 

Therefore the monodromy inside ${\mathcal L}_3$ around the $\Z_5$ point is given by 
\begin{equation}\label{e:m2}
\begin{split}
\ms{M}_{\Z_5}\, &= \, \ms{T}_{i_*\O_{D_4}}\comp \, \ms{M}_{\Z_2}\\
\, &= \, \ms{T}_{i_*\O_{D_4}} \comp \,  \ms{T}_{j_*\O_{D_5}(-f)} \, \comp   \, \ms{T}_{j_*\O_{D_5}}  \comp \, \ms{L}_{D_2}\,.
\end{split}
\end{equation}

\subsection{Quantum symmetries}		\label{s:M22}

It is an interesting question to ask how the $\Z_2$ (resp. $\Z_3$) quantum symmetry of the partially resolved $\Z_2$ (resp. $\Z_3$) orbifold is realized in the derived category setup. Accordingly, we would like to compute the action of $(\ms{M}_{\Z_2})^2$ on a generic object. Unfortunately this seems too hard at this moment.\footnote{The analogous statement for $\C^2/\Z_3$  was proved in \cite{en:fracC2}, while the compact case has been analyzed in \cite{en:Alberto}.} Nevertheless, we can compute the Chern character of $(\ms{M}_{\Z_2})^2$ acting on a generic K-theory class $x\in K(X)$. 

Inspecting the form of $\ms{M}_{\Z_2}$ in Eq.~(\ref{e:m1}) we see that in order to compute $\ch((\ms{M}_{\Z_2})^2 x)$  some general properties might be of use. Taking the Chern character of both sides in Eq.~(\ref{e:refl}) one obtains \cite{ST:braid,navigation}:
\begin{equation}\label{e:monch}
\ch\big(\ms{T}_{\mathsf{A}}  (\mathsf{B})\big)
=\ch(\mathsf{B})-\langle\mathsf{A},\mathsf{B}\rangle\ch(\mathsf{A})\,,
\end{equation}
where $\langle\mathsf{A},\mathsf{B}\rangle$ is an Euler characteristic:
\begin{equation}
\langle\mathsf{A},\mathsf{B}\rangle =
\sum_i (-1)^i\dim\Ext_{\D(X)}^i(\mathsf{A},\mathsf{B})\,.
\end{equation}
The Grothendieck--Riemann--Roch theorem gives a useful way to compute this:
\begin{equation}\label{e:monch1}
\langle\mathsf{A},\mathsf{B}\rangle = \int_X \ch(\mathsf{A}^{\!\vee}) \ch(\mathsf{B})\td(X)\,.
\end{equation}

Using (\ref{e:monch}) and (\ref{e:monch1}) one can show that for any K-theory class $x$
\begin{equation}\label{e:moni1}
\ch((\ms{M}_{\Z_2})^2\, x)= e^{D_1}\, \ch(x)\,
\end{equation}
and
\begin{equation}\label{e:moni2}
\ch((\ms{M}_{\Z_3})^3\, x)= e^{D_2}\, \ch(x)\,.
\end{equation}
This suggests that both $(\ms{M}_{\Z_2})^2$ and  $(\ms{M}_{\Z_3})^3$ act like large radius monodromies. The result is  surprising at first, but a similar fact has been observed before \cite{navigation,en:fracC2}, and it is consistent with the general statement that monodromy at an orbifold point has to be associated with $B$-field components other than the blow-up mode of the orbifold. For more on this we refer to \cite{navigation,en:fracC2}.

A similarly statement is true at the $\Z_5$ orbifold point:
\begin{equation}\label{e:moni3}
\ch((\ms{M}_{\Z_5})^5\, x)= \, \ch(x)\,.
\end{equation}
We note that $\ch((\ms{M}_{\Z_2})\, x)$ does  not have a simple expression, and similarly for $\ch((\ms{M}_{\Z_3})\, x)$ and $\ch((\ms{M}_{\Z_3})^2\, x)$, and for the lower powers of $\ms{M}_{\Z_5}$.

\section{The $\C^3/\Z_5$ fractional branes}	\label{s:cnfract}

In this section we use the $\Z_5$ monodromy action found in the previous section to generate a collection of fractional branes, and study some of their properties. As a starting point we need to know one of the fractional branes. We assume that the D5-brane wrapping the exceptional divisor $D_5$ is one of the fractional branes. This is a natural  assumption as long as we do not make any claims about the rest of the fractional branes. It is reasonable to expect that by various monodromy transformations any one of the fractional branes can be brought to this form. Instead of guessing the other two fractional branes, we look at the orbit of this D5-brane under the $\Z_5$ monodromy action. In the quiver language the fractional branes are the simple representations of the quiver, and are mapped into each other under the $\Z_5$ quantum symmetry. Therefore, the  fractional branes will necessarily form a length five orbit of the $\Z_5$ monodromy, which is an incarnation of the $\Z_5$ quantum symmetry.

\subsection{Generating fractional branes}

We start by recalling Eq.~(\ref{e:m2}), which gives us the form of the $\Z_5$ monodromy $\ms{M}_{\Z_5}$.
By the assumption made above the $1$st fractional brane is $j_*\O_{D_5}$. The others  are  $\ms{M}_{\Z_5}^k(j_*\O_{D_5})$, for $k=1,\ldots 4$. We start out by computing $\ms{M}_{\Z_5}(j_*\O_{D_5})$.

\subsubsection{Computing $\ms{M}_{\Z_5}(j_*\O_{D_5})$}

The first step is quite trivial:
\begin{equation}
\xymatrix{j_*\O_{D_5} \ar[rr]^{\ms{L}_{D_2}} && j_*\O_{D_5}(s+3f)}.
\end{equation}
We act on this with $\ms{T}_{\O_{D_5}}$,\footnote{To simplify the notation we denote $\ms{T}_{i_*\O_{D_k}}$ by $\ms{T}_{D_k}$.} and use the fact that $\R j_*$ is a triangulated functor, to obtain
\begin{equation}
\xymatrix{j_*\O_{D_5}(s+3f) \ar[rr]^-{\ms{T}_{\O_{D_5}}} && 
\Cone { j_*\O_{D_5}^{\oplus 5} \longrightarrow j_*\O_{D_5}(s+3f)}	}.
\end{equation}
The intermediate steps above involved using the spectral sequence (\ref{SS1}), but we will always suppress the details. In fact almost every step in this section involves one or more spectral sequences.

Applying $\ms{T}_{j_*\O_{D_5}(-f)}$ to this gives something interesting:
\begin{equation}
C=
\Cone {  j_*\O_{D_5}(-f)^{\oplus 3}[1] \longrightarrow \Cone {  j_*\O_{D_5}^{\oplus 5} \rightarrow j_*\O_{D_5}(s+3f)} }.
\end{equation}
We call this last complex $C$. It is easy to show that 
\begin{equation}
\ch (C) = -\ch(j_*\O_{D_5}(-s))\,,
\end{equation}
moreover, by naive counting $C$ is a line bundle on $D_5$. Therefore, it is natural to suspect the following 
\begin{lemma}\label{ex2}
$C \iso j_*\O_{D_5}(-s)[1] \,.$
\end{lemma}

\begin{proof}
Since the functor $\R j_*$ preserves triangles, and the morphisms in $C$ come from $\Hom^0_{D_5}$ (rather than higher $\Hom$'s on $D_5=\F_3$, which would be transformed by the  spectral sequence into $\Hom^0_X$)\footnote{This follows from the spectral sequence computation.}, we have that 
\begin{equation}
\begin{split}
C=&
\Cone {  j_*\O_{D_5}(-f)^{\oplus 3}[1] \longrightarrow \Cone {  j_*\O_{D_5}^{\oplus 5} \rightarrow j_*\O_{D_5}(s+3f)} }\\
&\iso  j_*\Cone {  \O_{D_5}(-f)^{\oplus 3}[1] \longrightarrow \Cone {  \O_{D_5}^{\oplus 5} \rightarrow \O_{D_5}(s+3f)} }\,.
\end{split}
\end{equation}
Therefore $C= j_* C_5$, where
\begin{equation}
C_5=
\Cone { \O_{D_5}(-f)^{\oplus 3}[1] \longrightarrow \Cone {  \O_{D_5}^{\oplus 5} \rightarrow \O_{D_5}(s+3f)} } \,.
\end{equation}
Using the long exact sequences associated to cones, one can show that 
\begin{equation}
\Hom_{D_5}^k (C_5,\O_{D_5}(-s)[1])  = \left\{ 
\begin{array}{rr}
\C & \mbox{for $k=0$}	\\
0   & \mbox{otherwise}.
\end{array}	\right.
\end{equation}
Therefore there is a map $f\!: C_5\to \O_{D_5}(-s)[1]$ in $\D(D_5)$. We will show that $\Cone{f}=0$ in $\D(D_5)$, and therefore $f$ is a quasi-isomorphism. 

To show that $\Cone{f}=0$ in $\D(D_5)$ it is sufficient to prove that 
\begin{equation}\label{e:vanHi}
\Hom_{D_5}^k (E_i, \Cone{f}) =0
\end{equation}
for a complete exceptional collection $E_i$ on  $D_5=\F_3$. This is a {\em spanning class} in Bridgeland's sense \cite{Bridgeland:FMequiv}. A convenient choice is the strong  exceptional collection 
\begin{equation}\label{e:F3expt}
\O\,,\; \O(f)\,,\; \O(s+3f)\,,\; \O(s+4f)\,,
\end{equation}
on $\F_3$ \cite{Rudakov:Book,Auroux}. Using this collection, it is a tedious but straightforward exercise to prove (\ref{e:vanHi}). 
\end{proof}

The final step of computing $\ms{M}_{\Z_5}(j_*\O_{D_5})$ involves using the spectral sequence (\ref{SS3}) and results in
\begin{equation}
\xymatrix{ j_*\O_{D_5}(-s)[1] \ar[rr]^-{\ms{T}_{\O_{D_4}}} && 
\Cone {i_*\O_{D_4} \longrightarrow j_*\O_{D_5}(-s)[1] }} = k_*\O_{D_4+D_5}[1]\,.
\end{equation}
The last equality follows from the exact triangle 
\begin{equation}\label{e:bpvdw}
\xymatrix{j_*\O_{D_5}(-s) \ar[r] & k_*\O_{D_4+D_5} \ar[r] &  i_*\O_{D_4} \ar[r] & j_*\O_{D_5}(-s)[1] },
\end{equation}
 where $k\!: D_4+D_5 \hookrightarrow X$ is the embedding map. 
The triangle stems from the short exact sequence:\footnote{Intuitively this formula is easy to understand. The inclusion map of $D$ into $C+D$ allows us to restrict functions from $C+D$ to $D$. This is the map $\O_{C+D} \rightarrow \O_{D}$. The kernel of this map consists of those functions on $C$ that vanish at the intersection point with $D$: $\O_{C}(-D)$. For a rigorous proof see \cite{barthpvdv}.}
\begin{equation}\label{e:bpvdw1}
\xymatrix{0 \ar[r] & \O_{C}(-D) \ar[r] & \O_{C+D} \ar[r] & \O_{D} \ar[r] & 0}\,.
\end{equation}

Thus the $2$nd fractional brane is $k_*\O_{D_4+D_5}[1]$, and is a D5-brane that wraps both exceptional divisors, $D_4$ and $D_5$. Eq.~(\ref{e:bpvdw}) also shows that the D5-brane $k_*\O_{D_4+D_5}$ wraps the intersection of $D_4$ and $D_5$ only once.

\subsubsection{Computing $(\ms{M}_{\Z_5})^2(j_*\O_{D_5})$}

To determine the $3$rd fractional brane we apply the $\Z_5$ monodromy again. Now the starting point is the second fractional brane  $k_*\O_{D_4+D_5}[1]$ from the previous section. Since all the operations that we perform are functors between triangulated categories, they all commute with the $[1]$ shift functor, and therefore 
\begin{equation}
\ms{M}_{\Z_5}(k_*\O_{D_4+D_5}[1]) = \ms{M}_{\Z_5}(k_*\O_{D_4+D_5})[1]\,.
\end{equation}
So it suffices to compute $\ms{M}_{\Z_5}(k_*\O_{D_4+D_5})$. First of all
\begin{equation}
\xymatrix{k_*\O_{D_4+D_5} \ar[rr]^-{\ms{L}_{D_2}} &&  k_*\O_{D_4+D_5}(D_2)\,.}
\end{equation}

For the next step we need $\RHom_X(j_*\O_{D_5},k_*\O_{D_4+D_5}(D_2))$. We will determine this in two different ways. The first method will use the cohomology long exact sequence associated to an exact triangle. The second method will use the spectral sequence (\ref{eq:dcx}). Although the first method is a priori more straightforward, the spectral sequence is much more efficient. The fact that the two methods give the same result provides a consistency check for our calculations.

The exact triangle (\ref{e:bpvdw}) implies that
\begin{equation}\label{e:bpvdw51}
\xymatrix{j_*\O_{D_5}(3f) \ar[r] & k_*\O_{D_4+D_5}(D_2) \ar[r] &  i_*\O_{D_4} \ar[r] & j_*\O_{D_5}(3f)[1] }\,,
\end{equation}
is an exact triangle as well. Applying the covariant functor $\Ext^i_X(j_*\O_{D_5},-)$ to the exact triangle (\ref{e:bpvdw51}) gives the  long exact sequence:
\begin{equation}\label{e:lesd}
\Ext^{i}_X(j_*\O_{D_5}, j_*\O_{D_5}(3f))\longrightarrow  
\Ext^i_X(j_*\O_{D_5}, k_*\O_{D_4+D_5}(D_2))\longrightarrow 
\Ext^i_X(j_*\O_{D_5}, i_*\O_{D_4} ) 
\end{equation}
The spectral sequence (\ref{SS3}) tells us that $\Ext^i_X(j_*\O_{D_5}, i_*\O_{D_4} )=\delta_{i,1}\C^2$, while using the spectral sequence  (\ref{SS1}) gives
\begin{equation}\label{e:lesd1}
\Ext^i_X(j_*\O_{D_5}, j_*\O_{D_5}(3f)) = \left\{ 
\begin{array}{rr}
\C^4 & \mbox{for $i=0$}	\\
\C^2 & \mbox{for $i=2$}	\\
0   & \mbox{otherwise}.
\end{array}	\right.
\end{equation}
Using these two facts, the long exact sequence  (\ref{e:lesd}) tells us that $\Ext^i_X(j_*\O_{D_5}, k_*\O_{D_4+D_5}(D_2))=\delta_{i,0}\C^4$.

The same result can be obtained much quicker, if we apply the spectral sequence (\ref{eq:dcx}) to our case. By Serre duality 
\begin{equation}
\Ext^i_X(j_*\O_{D_5}, k_*\O_{D_4+D_5}(D_2))=\Ext^{3-i}_X(k_*\O_{D_4+D_5}(D_2),j_*\O_{D_5})
\end{equation}
The spectral sequence (\ref{eq:dcx}) then reads
\begin{equation}
\begin{xy}
\xymatrix@C=2mm{
   & \H^p (\F_3,  \O(-2s-8f)) & \\
   & \H^p (\F_3,  \O(-s-3f)) & }
\save="x"!LD+<-3mm,0pt>;"x"!RD+<0pt,0pt>**\dir{-}?>*\dir{>}\restore
\save="x"!LD+<0pt,-3mm>;"x"!LU+<0pt,-2mm>**\dir{-}?>*\dir{>}\restore
\save!CD+<0mm,-4mm>*{p}\restore
\save!CL+<-3mm,0mm>*{q}\restore
\end{xy}
\begin{xy}
\xymatrix@C=10mm{
    &  &\\
    &= &\\
     &&\\}
\end{xy}
\begin{xy}
\xymatrix@C=15mm{
    0 &    0 & \C^4 \\
    0 &    0 & 0 }
\save="x"!LD+<-3mm,0pt>;"x"!RD+<0pt,0pt>**\dir{-}?>*\dir{>}\restore
\save="x"!LD+<0pt,-3mm>;"x"!LU+<0pt,-2mm>**\dir{-}?>*\dir{>}\restore
\save!CD+<0mm,-4mm>*{p}\restore
\save!CL+<-3mm,0mm>*{q}\restore
\end{xy}
\end{equation}
and therefore $\Ext^i_X(j_*\O_{D_5}, k_*\O_{D_4+D_5}(D_2))=\delta_{i,0}\C^4$, as we saw before.

Therefore we established that 
\begin{equation}
\xymatrix{ k_*\O_{D_4+D_5}(D_2) \ar[rr]^-{\ms{T}_{\O_{D_5}}} && 
\Cone{j_*\O_{D_5}^{\oplus 4}\to k_*\O_{D_4+D_5}(D_2)}\,.}
\end{equation}

Applying $\ms{T}_{j_*\O_{D_5}(-f)}$ to this gives:
\begin{equation}\label{e:befoct}
\begin{split}
&\xymatrix{ k_*\O_{D_4+D_5}(D_2) \ar[rrr]^-{\ms{T}_{\O_{D_5}(-f)}} &&&} \\
&~~\quad\xymatrix{\Cone{j_*\O_{D_5}(-f)^{\oplus 3}[1]\longrightarrow 
\Cone{j_*\O_{D_5}^{\oplus 4}\to k_*\O_{D_4+D_5}(D_2)}}\,.}
\end{split}
\end{equation}

The way it stands, this expression is not too revealing, but fortunately we can simplify it dramatically using a deep property of triangulated categories, known as the Octahedral Axiom.

\paragraph{The Octahedral Axiom}

Assume that we are given two distinguished triangles having an  object $B$ in common, $(A,B,C,i,j,k)$ and $(B,D,E,u,v,w)$ (these are the solid arrows):
\begin{equation}\label{e:xx}
\xymatrix@C=13mm{
&B\ar[lddd]^j\ar[rd]^u&\\
D\ar@{..>}[dd]|{[1]}_(0.4)v\ar[ur]|{[1]}^(0.4)w&&E
        \ar[ll]^v\\
&&\\
C\ar[rr]|{[1]}_(0.4)k&&A\ar[uuul]^i\ar@{..>}[uu]
}
\qquad\qquad\qquad
\xymatrix@C=13mm{
&B\ar[lddd]\ar[rd]&\\
D\ar@{.>}[dd]|{[1]}\ar[ur]|{[1]}&&E
        \ar[ll]\ar@{-->}[dddl]\\
&&\\
C\ar[rr]|{[1]}\ar@{-->}[dr]&&A\ar[uuul]\ar@{.>}[uu]\\
&{F}\ar@{-->}[ur]|{[1]}\ar@{-->}[uuul]&}
\end{equation}
This setting gives us automatically two maps: $u \,\comp i\!:A\to E$ and $j[1]\comp w\!: D\to C[1]$ --- these are the dotted lines above. Using the two maps we can construct two distinguished triangles: $(A,E,F_1,u\comp i,.,.)$ and $(C,F_2,D,.,.,j[1]\comp w)$. The Octahedral Axiom  states that $F_1=F_2$, and the newly created two extra faces of the octahedron commute.\footnote{Although this way of stating the octahedral axiom is not the standard one, it is equivalent to the more customary one \cite{GM:Hom}.} 

Altogether, four of the faces of the octahedron are distinguished triangles and the other four faces commute. We can present the upper an lower halves of the octahedron separately:

\def\octUp#1#2#3#4#5{
\xymatrix@C=9mm{
#1\ar[dd]\ar[dr]|{[1]}&\ar@{}|{\Delta}[d]&#4\ar[ll]\\
\ar@{}[r]|{\boxdot}&#5\ar[ur]\ar[dl]\ar@{}[r]|{\boxdot}&\\
#2\ar[rr]|{[1]}&\ar@{}|{\Delta}[u]&#3\ar[uu]\ar[ul]}}

\def\octDown#1#2#3#4#5{
\xymatrix@C=9mm{
#1\ar@{.>}[dd]|{[1]} &\ar@{}[d]|{\boxdot}&#4\ar[ll]\ar[dl]\\
\ar@{}|*+[o][F.]{\mbox{\scriptsize $\Delta$}}[r]&
#5\ar@{.>}[ul]\ar[dr]|{[1]}\ar@{}|{\Delta}[r]\ar@{}[d]|{\boxdot}&\\
#2\ar[rr]\ar@{.>}[ur]&&#3\ar[uu]}}

\begin{equation}
\octUp DCAEB\qquad\qquad\qquad\quad
\octDown DCAEF
\end{equation}
We signaled a distinguished triangle with the symbol $\Delta$, and a commuting triangle with $\boxdot$. 

Forgetting for a moment about the maps involved, the  octahedral axiom can be rewritten as
\begin{equation}\label{a:octhadral-1}
\Cone{C \longrightarrow\Cone{A\rightarrow E}} = \Cone{\Cone{C[-1]\rightarrow A}\longrightarrow E}\,.
\end{equation}
This comes from the observation that $D$ can be written in two different ways in terms of the other objects.


Now we can return to the monodromy computation. Using the octahedral axiom (\ref{a:octhadral-1}), the complex in (\ref{e:befoct}) becomes
\begin{equation}\label{a:octhbadr-11}
\xymatrix{\Cone{\Cone{j_*\O_{D_5}(-f)^{\oplus 3}\to j_*\O_{D_5}^{\oplus 4}} 
\longrightarrow k_*\O_{D_4+D_5}(D_2)}\,.}
\end{equation}
To proceed we recall from Eq.~(\ref{e:bpvdw51}) that
\begin{equation}\label{a:octhadr-11}
k_*\O_{D_4+D_5}(D_2)=\Cone{i_*\O_{D_4}[-1]\to j_*\O_{D_5}(3f)}\,,
\end{equation}
and use another form of the octahedral axiom (once again $D$ is written in two different ways):
\begin{equation}\label{a:octhadral-11}
\Cone{\Cone{A \to B} \longrightarrow F} = \Cone{B \longrightarrow \Cone{F[-1] \to A}}\,.
\end{equation}
Using (\ref{a:octhadral-11}) and (\ref{a:octhadr-11}), Eq.~(\ref{a:octhbadr-11}) becomes\footnote{Here $F=i_*\O_{D_4}$, $A=j_*\O_{D_5}(3f)$ and $B=\Cone{j_*\O_{D_5}(-f)^{\oplus 3}\to j_*\O_{D_5}^{\oplus 4}}$.}
\begin{equation}\label{a:bocthadral-11}
\Cone{\Cone{j_*\O_{D_5}(3f) \to \Cone{j_*\O_{D_5}(-f)^{\oplus 3}\to j_*\O_{D_5}^{\oplus 4}}} 
\longrightarrow i_*\O_{D_4}}.
\end{equation}
This expression is in fact much simpler than it looks, due to the following
\begin{lemma}
$C=
\Cone{j_*\O_{D_5}(3f) \to \Cone{j_*\O_{D_5}(-f)^{\oplus 3}\to j_*\O_{D_5}^{\oplus 4}}}
\iso 0 \,.$
\end{lemma}

\begin{proof}
The proof is identical in spirit to one from the previous section, and therefore we are going to be sketchy. The functor $j_*$ preserves triangles, and the morphisms in $C$ are $\Hom^0$'s on $D_5=\F_3$, so $C= j_* C_5$, where
\begin{equation}
C_5=\Cone { \O_{D_5}(3f) \to \Cone{\O_{D_5}(-f)^{\oplus 3}\to \O_{D_5}^{\oplus 4}} }\,.
\end{equation}
An explicit computation then shows that $\Hom_{D_5}^k (E_i, C_5) =0$ for the spanning class (\ref{e:F3expt}).
\end{proof}

Using the lemma we have that 
\begin{equation}
\xymatrix{ k_*\O_{D_4+D_5}(D_2) \ar[rrr]^-{\ms{T}_{\O_{D_5}(-f)}} &&& j_*\O_{D_4}\,.}
\end{equation}

Therefore the last step of the computation involves $\ms{T}_{\O_{D_4}}(i_*\O_{D_4})$. Here we can use a more general result\footnote{For a proof the reader can consult Lemma~4.1 in \cite{en:fracC2}.}
\begin{lemma}\label{lemma:41}
If $\mathsf{A}$ is an $n$-spherical object, then $\ms{T}_{\mathsf{A}} (\mathsf{A}) = \mathsf{A}[1-n]$.
\end{lemma}

For $n=3$ and $\mathsf{A}=i_*\O_{D_4}$\footnote{Prop.~3.15 of \cite{ST:braid} guarantees that $i_*\O_{D_4}$ is 3-spherical in the sense of Definition~\ref{def:spherical}, or one can check this directly using the spectral sequence (\ref{SS1}).}
the lemma gives $\ms{T}_{i_*\O_{D_4}}(i_*\O_{D_4})= i_*\O_{D_4}[-2]$, and thus 
\begin{equation}
\ms{M}_{\Z_5}(k_*\O_{D_4+D_5}[1]) = i_*\O_{D_4}[-1]\,.
\end{equation}
This establishes $i_*\O_{D_4}[-1]$ as the $3$rd fractional brane.

\subsubsection{Computing $(\ms{M}_{\Z_5})^3(j_*\O_{D_5})$}

The first step is totally trivial since $D_2$ and $D_4$ are disjoint:
\begin{equation}
\xymatrix{i_*\O_{D_4} \ar[rr]^{\ms{L}_{D_2}} && i_*\O_{D_4}}\,.
\end{equation}
Now we act on this with $\ms{T}_{\O_{D_5}}$, and use the spectral sequence (\ref{SS1}):
\begin{equation}
\xymatrix{i_*\O_{D_4}  \ar[rr]^-{\ms{T}_{\O_{D_5}}} && 
\Cone { j_*\O_{D_5}^{\oplus 2}[-1] \longrightarrow i_*\O_{D_4}}	}\,.
\end{equation}
Applying $\ms{T}_{j_*\O_{D_5}(-f)}$ gives:
\begin{equation}
\Cone {  j_*\O_{D_5}(-f) \longrightarrow \Cone {  j_*\O_{D_5}^{\oplus 2}[-1] \rightarrow i_*\O_{D_4}} }\,.
\end{equation}
We can rewrite this expression using the octahedral axiom (\ref{a:octhadral-1}):
\begin{equation}\label{ex1}
\Cone {  \Cone { j_*\O_{D_5}(-f) \longrightarrow  j_*\O_{D_5}^{\oplus 2}}[-1] \rightarrow i_*\O_{D_4} }\,.
\end{equation}
The first cone simplifies if we use the Koszul resolution\footnote{The maps are the correct ones needed for this to work.} (for the intersection of two fibers on $D_5=\F_3$, which is  empty, and recall that $\O_\emptyset=0$):
\begin{equation}
\ses{\O_{D_5}(-2f)}{\O_{D_5}(-f)^{\oplus 2}}{\O_{D_5}}\,.
\end{equation}
This shows that 
\begin{equation}
\Cone { j_*\O_{D_5}(-f) \longrightarrow  j_*\O_{D_5}^{\oplus 2}}= \O_{D_5}(f)\,,
\end{equation}
and therefore (\ref{ex1}) becomes
\begin{equation}
\Cone { \O_{D_5}(f) [-1] \rightarrow i_*\O_{D_4} }\,.
\end{equation}
Finally
\begin{equation}
\xymatrix{ \Cone { \O_{D_5}(f) [-1] \to i_*\O_{D_4}}  \ar[rr]^-{\ms{T}_{\O_{D_4}}} && 
\Cone {i_*\O_{D_4} \to \Cone { \O_{D_5}(f) [-1] \to i_*\O_{D_4}} } 
\iso \O_{D_5}(f)	}
\end{equation}
In other words, the  $4$th fractional brane, $\O_{D_5}(f)$, is a 5-brane  wrapping the $\F_3$ with D3-brane flux turned on.

\subsubsection{Computing $(\ms{M}_{\Z_5})^4(j_*\O_{D_5})$}	\label{s:3rdtime}

First:
\begin{equation}
\xymatrix{j_*\O_{D_5}(f) \ar[rr]^{\ms{L}_{D_2}} && j_*\O_{D_5}(s+4f)}\,.
\end{equation}
Acting with $\ms{T}_{\O_{D_5}}$ one obtains
\begin{equation}
\xymatrix{j_*\O_{D_5}(s+4f) \ar[rr]^-{\ms{T}_{\O_{D_5}}} && 
\Cone { j_*\O_{D_5}^{\oplus 7} \longrightarrow j_*\O_{D_5}(s+4f)}	}\,.
\end{equation}

Applying $\ms{T}_{j_*\O_{D_5}(-f)}$   gives 
\begin{equation}
C=
\Cone {  j_*\O_{D_5}(-f)^{\oplus 5}[1] \longrightarrow \Cone {  j_*\O_{D_5}^{\oplus 7} \rightarrow j_*\O_{D_5}(s+4f)} }\,.
\end{equation}
We named this last complex $C$. It is easy to show that 
\begin{equation}
\ch (C) = -\ch(j_*\O_{D_5}(f-s))\,,
\end{equation}
moreover, by naive counting $C$ is a line bundle on $D_5$. Therefore, it is natural to suspect that
\begin{lemma}
$C \iso j_*\O_{D_5}(f-s)[1] \,.$
\end{lemma}

\begin{proof}
The proof is by now standard: the crucial point is once again the fact that every map in $C$ is a $\Hom^0$ on $D_5=\F_3$, and thus $C= j_* C_5$, where
\begin{equation}
C_5=
\Cone { \O_{D_5}(-f)^{\oplus 5}[1] \longrightarrow \Cone {  \O_{D_5}^{\oplus 7} \rightarrow \O_{D_5}(s+4f)} } \,.
\end{equation}
Using the long exact sequences associated to the cones one  shows that 
\begin{equation}
\Hom_{D_5}^k (C_5,\O_{D_5}(f-s)[1])  = \left\{ 
\begin{array}{rr}
\C & \mbox{for $k=0$}	\\
0   & \mbox{otherwise}.
\end{array}	\right.
\end{equation}
Therefore there is a map $f\!: C_5\to \O_{D_5}(-s)[1]$ in $\D(D_5)$, and it turns out that $\Cone{f}$ is orthogonal to the spanning class (\ref{e:F3expt}), and we can conclude like in Lemma~\ref{ex2}.
\end{proof}

Using the lemma the final step of $\ms{M}_{\Z_5}(j_*\O_{D_5}(f) )$ is
\begin{equation}
\ms{T}_{\O_{D_4}}( j_*\O_{D_5}(f-s)[1])=
\Cone {i_*\O_{D_4}^{\oplus 2} \longrightarrow j_*\O_{D_5}(f-s)[1] }\,.
\end{equation}
This is another  5-brane  wrapping both $D_4$ and $D_5$. One expects these branes in order to be able to describe the decays of all the $\C^3/\Z_3$ fractional branes (which can be chosen to be $\{\Omega_{\P^2}^k(k)[k]\}_{k=0}^2$).

\subsection{The quiver}

Let us summarize the lengthy computation of this section. We have shown the following mappings under the action of $\ms{M}_{\Z_5}$:
\begin{equation}
\begin{split}
&j_*\O_{D_5} \mapsto k_*\O_{D_4+D_5}[1],\qquad  j_*\O_{D_4+D_5} \mapsto i_*\O_{D_4}[-2],\\ 
&i_*\O_{D_4}\mapsto j_*\O_{D_5}(f),\qquad  
j_*\O_{D_5}(f)\mapsto \Cone {i_*\O_{D_4}^{\oplus 2}\to j_*\O_{D_5}(f-s)[1] }.
\end{split}
\end{equation} 
Shifting appropriately, the following list of five objects forms an orbit of the $\Z_5$ orbifold quantum symmetry:
\begin{equation}\label{efracColl}
j_*\O_{D_5}[1],\; k_*\O_{D_4+D_5}[2],\; i_*\O_{D_4},\; j_*\O_{D_5}(f),\; \Cone {i_*\O_{D_4}^{\oplus 2}\to j_*\O_{D_5}(f-s)[1] }.
\end{equation} 
Therefore we have a potential sets of fractional branes. Let us call them $\{\c F_i \}_{i=1}^5$. All of them are automatically $3$-spherical since they were obtained by autoequivalence, and the initial one, $j_*\O_{D_5}$, is $3$-spherical.

The first thing we need to check is whether their charges, measured by K-theory, add up to that of the D3-brane. One easily computes the Chern characters, and has that 
\begin{equation}
\sum_{i=1}^5 \ch (\c F_i )= \ch (\O_{pt})\,.
\end{equation} 

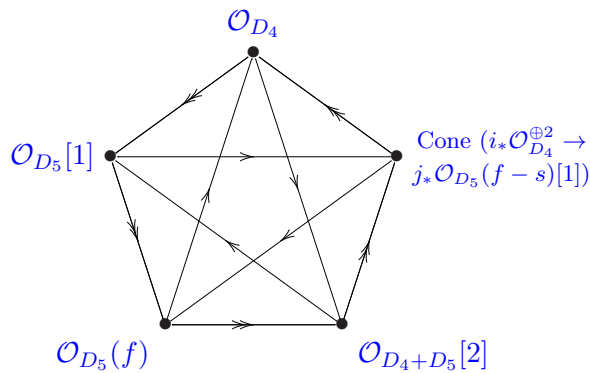
\begin{figure}[h]
\begin{equation}\nonumber
\begin{xy} <20mm,0cm>:
,{\xypolygon5"A"{{\bullet}}}
,"A2"*++!D{\color{blue}\O_{D_4}}
,"A3"*++!R{\color{blue}\O_{D_5}[1]}
,"A4"*++!UR{\color{blue}\O_{D_5}(f)}
,"A5"*++!UL{\color{blue}\O_{D_4+D_5}[2]}
,"A1"*++!L\txt{\color{blue}\footnotesize Cone $(i_*\O_{D_4}^{\oplus 2}\to$ \\\color{blue}\footnotesize$ j_*\O_{D_5}(f-s)[1]) $}
\ar@{-}^(0.5)*\dir{>>} "A1"; "A2"
\ar@{-}^(0.5)*\dir{>>} "A2"; "A3"
\ar@{-}^(0.5)*\dir{>>} "A3"; "A4"
\ar@{-}^(0.5)*\dir{>>} "A4"; "A5"
\ar@{-}^(0.5)*\dir{>>} "A5"; "A1"
\ar@{-}^(0.5)*\dir{>} "A3"; "A1"
\ar@{-}^(0.5)*\dir{>} "A4"; "A2"
\ar@{-}^(0.5)*\dir{>} "A5"; "A3"
\ar@{-}^(0.5)*\dir{>} "A1"; "A4"
\ar@{-}^(0.5)*\dir{>} "A2"; "A5"
\end{xy}
\end{equation}
  \caption{The  $\C^3/\Z_5$ quiver.}
  \label{f:c2z3}
\end{figure}

Next we compute the $\Ext^1$-quiver of the collection (\ref{efracColl}). Using the spectral sequences from Appendix~\ref{s:ss} we obtain the well-known $\C^3/\Z_5$  quiver, depicted in Fig.~\ref{f:c2z3}.

Moreover, the $\Ext^0$'s and $\Ext^3$'s  between different branes are all zero. Therefore the potentially tachyonic strings are missing. For every $\Ext^1$ there will be an $\Ext^2$ by Serre duality, and the $\Ext^2$'s will give the correct superpotential. 

Finally we need to check that the central charge, and hence the mass, of the $\c F_i$'s is {\em a fifth} of the D3-brane  central charge {\em at the} $\Z_5$ point in moduli space. This can be done using the expression for the central charges in terms of the periods given in \cite{DelaOssa:2001xk}. This works because the  central charges are determined by the large volume asymptotics, which depend only on the Chern character of the brane.

\subsection{Connection with the McKay correspondence}

The classical version of the McKay correspondence \cite{McKay} relates the representations of a finite subgroup  $\Gamma$  of $\SL(2,\C)$ to the cohomology of the minimal resolution of the Kleinian singularity $\C^2/\Gamma$. 

A solid understanding of the McKay correspondence culminated with the work of Bridgeland, King and Reid (BKR) \cite{Mukai:McKay}, who showed that in dimensions two and three the McKay correspondence is an equivalence of two very different derived categories.\footnote{The BKR proof generalizes to higher dimensions, provided there exists a crepant resolution \cite{Mukai:McKay}.} 
Let us review their construction applied to  $\C^3/\Z_5$. As we will see, the McKay correspondence provides a set of fractional branes, which is different from what we obtained by monodromy.

First we have the covering map $q\!: \C^3\longrightarrow \C^3/\Z_5$, and the map $\tilde{p}\!: X\longrightarrow \C^3/\Z_5$ corresponding to the resolution of singularities. Using these two maps we can consider the fiber product $Y$ of $\C^3$ and $X$ over $\C^3/\Z_5$:
\begin{equation}
\xymatrix@C=20mm{
Y \ar[r]^{q_2} \ar[d]_{q_1} & \C^3 \ar[d]^{q}  \\
X \ar[r] ^{\tilde{p}} & \C^3/\Z_5	}
\end{equation}

Let $\Coh^{\Z_5} (\C^3)$ be the category of $\Z_5$-equivariant coherent sheaves on $\C^3$. BKR show that the functor 
\begin{equation}\label{eq:psiphi}
\Phi = \R q_{2*}\comp q _{1}^*: \D(X)\longrightarrow \D(\Coh^{\Z_5} (\C^3))   
\end{equation}
is an  equivalence of categories. This statement implies the classical McKay correspondence.

There is another equivalence at hand, although this one is more mundane: the one-to-one correspondence between the representations of the $\C^3/\Z_5$ McKay quiver and  the category of $\Z_5$-equivariant coherent sheaves on $\C^3$ (for a review  see \cite{Paul:TASI2003}). In the language of quiver representations the fractional branes are the simple objects, i.e. with no sub-objects; the representations with all but one node assigned the trivial vector space, and all arrows are assigned the $0$ morphisms. We can compose this equivalence with $\Phi$ from (\ref{eq:psiphi}) and ask for the inverse images of the simples in $\D(X)$. This  provides a set of fractional branes. Unfortunately this question has not been answered yet in the literature. So far the   answer is known only for $\C^n/\Z_n$ (for $n=3$ see, e.g., \cite{Paul:TASI2003}) and $\C^2/\Gamma$ \cite{Kapranov:Vasserot}.

If we simplify the question and ask {\em only} about K-theoretic inverse images, then one can use the technology of Ito and Nakajima \cite{Ito:Nakajima}. This process was carried out in \cite{DelaOssa:2001xk}, with the following answer:
\begin{equation}\label{e:colMcKay}
j_* \O_{D_5}(-s) + i_*\O_{D_4}, \, -j_* \O_{D_5}(-s-f) -i_*V, \, i_*\O_{D_4}(-1),\,  -j_*\O_{D_5}(-s), \, j_*\O_{D_5}(-s-f).
\end{equation} 
Where $V$ is a rank $2$ bundle on $\P^2$, with $\ch V =  2-H -H^2/2$; $H$ being the hyperplane class on $\P^2$. 

This collection is to be contrasted with the one obtained in Section~\ref{s:cnfract}, more precisely  (\ref{efracColl}). Next we  elucidate their connection using Seiberg duality. 

\subsubsection{Seiberg duality}

The original Seiberg duality \cite{Seiberg:1994pq} is a low-energy equivalence between $N = 1$ supersymmetric gauge theories: an $SU(N_c)$  theory with $N_f$ fundamental flavors  and no superpotential, and an $SU(N_f -  N_c)$ theory with $N_f$ fundamental  magnetic  flavors with a superpotential containing mesons. The duality says that both flow to the same theory in the infrared. We will use the Berenstein-Douglas \cite{Berenstein:2002fi} extension of Seiberg duality, which has a natural stratification. In its simplest form it amounts to a base change for the branes. Since the new basis usually involves anti-branes in the language of the old basis, this change is most naturally done in the derived category of coherent sheaves, rather than sheaves alone. Therefore in this form  Seiberg duality is an autoequivalence of the derived category of coherent sheaves, which by Orlov's theorem  (Theorem \ref{thm:orlov}) is a Fourier-Mukai functor. 

The most general form of Seiberg duality arises when the {\em $t$-structure} of the derived category is changed. This is usually achieved by the use of tilting complexes \cite{Berenstein:2002fi}. What makes this possible is the underlying fact that there are different abelian categories with equivalent derived categories. 

Thus, in general, the difference between two collections of fractional branes can only be partially attributed to a choice of basepoint, since tiltings are more general than auto-equivalences. The McKay collection, although not explicitly, but inherently is associated to the vicinity of the orbifold point. The collection obtained by monodromies explicitly involved the choice of a basepoint for the loops in the moduli space, and this basepoint was in the vicinity of the  large volume point. Therefore it is reasonable to expect that the two collections differ only by a change in basepoint.

Indeed, one finds that the two collections (\ref{e:colMcKay}) and  (\ref{efracColl}) are related by monodromy around a point in moduli space where the brane wrapping both $D_4$ and $D_5$ is becoming massless, in other words $\ms{T}_{\O_{D_4+D_5}}$. The intuition for this fact comes from two observations. First, one recognizes that in K-theory $j_* \O_{D_5}(-s) + i_*\O_{D_4}=k_*\O_{D_4+D_5}$, and second, that $\ch V= \ch\Omega_{\P^2}(1)$, where $\Omega_{\P^2}$ is the cotangent bundle of $\P^2$. The authors of \cite{DelaOssa:2001xk} seem to be unaware of these facts.

Since $\O_{D_4+D_5}$ is 3-spherical\footnote{It was obtained from a 3-spherical object, $j_*\O_{D_5}$, by an autoequivalence.}, Lemma~\ref{lemma:41} tells us that
\begin{equation}
\ms{T}_{\O_{D_4+D_5}}(k_*\O_{D_4+D_5}[2])=k_*\O_{D_4+D_5}[2-2]=k_*\O_{D_4+D_5}.
\end{equation}
This establishes a link between the two sets. One can go further, and show that $\ch(\ms{T}_{\O_{D_4+D_5}} (\c F_i))$, for $\c F_i$ from the collection (\ref{efracColl}), is precisely the set given in (\ref{e:colMcKay}). This is the most one can say at this point, since the collection (\ref{e:colMcKay}) is defined only in K-theory.

One can turn the argument around, and propose that the above relationship lifts to $\D(X)$, in other words the McKay collection is given by $\ms{T}_{\O_{D_4+D_5}}$ applied to (\ref{efracColl}). This is easy to compute, since all the necessary $\Hom$'s can be read off from the quiver. Applying $\ms{T}_{\O_{D_4+D_5}}$  to (\ref{efracColl}) yields
\begin{equation}\label{e:colMcKay23}
i_*\O_{D_4}(-1)[2],\,  k_*\O_{D_4+D_5}, \, j_*\O_{D_5}(-s)[1], \, \Cone{k_*\O_{D_4+D_5}^{\ \oplus 2}\!\! \to\! j_*\O_{D_5}(f)}, \,  j_*\O_{D_5}(-s-f)[2].
\end{equation} 
Every step above is straightforward except for the last term, $ j_*\O_{D_5}(-s-f)[2]$, where one needs to use the octahedral axiom (\ref{a:octhadral-1}). 

It would be interesting to reproduce  the collection (\ref{e:colMcKay23}) by a direct computation, using $\Phi$ from the McKay correspondence (\ref{eq:psiphi}), as done for   $\C^3/\Z_3$ in \cite{Eric:DC2}, and for $\C^2/\Z_n$ in \cite{Kapranov:Vasserot}.

\section{The $\C^n/\Z_m$ quiver from partial resolutions}

The partial resolutions of the $\C^n/\Z_m$ singularity form a partially ordered set. The simplest partial resolutions involve blowing up {\em only one} exceptional divisor. This is particularly easy to do torically. For $\C^3/\Z_5$ we sketched the partial resolutions in Fig.~\ref{fig:4triang}. In fact there is a one-to-one correspondence between the different $\Z_m$ actions on $\C^n$, and the different compact divisors in the resolution of the $\C^n/\Z_m$ singularity. We outline this relationship briefly, as it plays a useful role in proving the main result of this section.

Since $\Z_m$ is cyclic, the action is totally specified by how a primitive generator acts on $\C^n$: 
\begin{equation}\label{e:crepact}
(z_1,\ldots,z_n)\mapsto (\omega_n^{a_1} z_1,\ldots,\omega_n^{a_n}  z_n)\,, \qquad (\omega_n) ^n=1 .
\end{equation}
We fix the $a_i$'s to be between $0$ and $m-1$.

From the string theory point of view the most interesting $\Z_m$ actions  are those where $\Z_m$ is a subgroup of $\SL(m,\C)$, since these admit crepant resolutions. From now on we make the following assumption
\begin{equation}\label{e:crep}
m=\sum_{i=1}^n a_i \,.
\end{equation} 

There is a convenient toric representation for the $\C^n/\Z_m$ variety. As explained in Section~2.2 of \cite{Fulton:T}, the toric fan consists of only one cone, generated by the unit vectors of $\Z^n$, while the $N$ lattice of the toric variety is:
\begin{equation}\label{e:latN}
N= \Z^n +\dfrac{1}{m}(a_1,\ldots, a_n) \Z\,.
\end{equation} 

Let $\{e_i\}_{i=1}^n$ be the unit vectors of $\Z^n$. Condition (\ref{e:crep}) guarantees that the vectors $e_i$ and $v=\frac{1}{m}(a_1,\ldots, a_n)$ all lie on the hyperplane $\sum_{i=1}^n x_i=1$. It is also clear that $v$ is an affine combination of the $e_i$'s. This representation links the different group actions (\ref{e:crepact})  to the various rational affine combinations of the base vectors $e_i$. 

For simplicity we assume that the  $a_i$'s do not have a common divisor. If this were not the case, than this common divisor $d$ would also divide $m=\sum  a_i$, and the $\Z_d$ subgroup of $\Z_m$ would act trivially, effectively reducing the action to $\Z_\frac{m}{d}$. Of course this statement is valid for schemes, and would fail if we viewed $\C^n/\Z_m$ as a smooth Deligne-Mumford stack. 

The lattice vector $v=\frac{1}{m}(a_1,\ldots, a_n)$ gives rise to a torus-invariant divisor $D_v$. As usual, the divisor itself is a toric variety, and  with our assumption on the $a_i$'s it is the weighted projective space $\P^{n-1}(a_1,\ldots, a_n)$. We sketch the argument.

First one observes that 
\begin{equation}\label{eqrel}
\sum_{i=1}^n a_i(e_i -v)=\sum_{i=1}^n a_i e_i -m v=0\,.
\end{equation} 
The $N$ lattice of $D_v$ by definition is $N_{D_v}= N/\Z  v$ \cite{Fulton:T}. The star of $v$ consists of every cone, and the rays of $D_v$ are $e_i -v$. Let us denote by $\bar{u}$ the class of the image of $u\in N$ under the natural projection $N \to N_{D_v}$. Therefore the relation (\ref{eqrel}) descends to
\begin{equation}\label{eqrel1}
\sum_{i=1}^n a_i\, \bar{ e_i }=0\,.
\end{equation} 
This is the ``signature'' relation for $\P^{n-1}(a_1,\ldots, a_n)$. One still needs to show the minimality of this relation. 
\begin{proof}
We assume that there is another relation $\sum_{i=1}^n r_i\bar{ e_i }=0$. Lifting this to $N$ means $\sum_{i=1}^n r_i { e_i }=\alpha v$, for some $\alpha\in \Z$. In other words 
\begin{equation}\label{eqrel2}
r_i=\dfrac{\alpha}{m}\, a_i\,,\qquad \mbox{for all $1\leq i \leq n$}. 
\end{equation}
Now for no $k$ is $a_k=0$ (otherwise we wouldn't have a $\Z_m$ action on $\C^n$ but on a lower dimensional space), and thus $r_k\neq 0$, i.e., the length of the relation is minimal. Furthermore, if $\alpha/m <1$, then $m/\alpha$ divides $a_i$ for all $1\leq i \leq n$. We already assumed that the  $a_i$'s do not have a common divisor, and thus $\alpha/m \geq 1$. Now summing both sides of (\ref{eqrel2}) gives $\sum_{i=1}^n r_i= \alpha\geq m$, and thus (\ref{eqrel1}) is indeed minimal.
\end{proof}

The McKay correspondence  gives an equivalence between quiver representations and sheaves on the resolved space, but it glosses over the partial resolutions. One can fill in the gap, by recasting it slightly into the language of stacks. First recall that there is an equivalence of categories between $\C^n/G$ quiver representations and coherent sheaves on the quotient stack  $[\C^n/G]$. Therefore the McKay correspondence reads as
\begin{equation}
\D([\C^n/G])\cong \D(\mbox{crepant resolution of $\C^n/G$})\,.
\end{equation}

Kawamata generalized the above statement, and for $G$ abelian he proved that \cite{Kawamata:GMcKay}:
\begin{equation}
\D([\C^n/G])\cong \D(\mbox{{\bf partial} crepant {\bf stacky} resolution})\cong \D(\mbox{crepant resolution})
\end{equation}
where in the middle one has to consider the partially resolved space as a stack. 

Therefore it makes sense to talk about fractional branes on the partially resolved space, and ask what they are. The strategy of this section is to use an appropriate set of objects on the exceptional divisor of the resolution to model the fractional branes. This strategy was successfully deployed in \cite{Herzog:2005sy} as well.

We already saw that the divisor $D_v$ is $\P^{n-1}(a_1,\ldots, a_n)$. In the light of Kawamata's work, we need to consider the stack ${\mathbf P}^{n-1}(a_1,\ldots, a_n)$ instead. The reason for this is intuitively clear: in order for $D_v$ to be able to capture the fact that it provides a  partial resolution for the $\C^n/\Z_m$ singularity, we have to retain more information than it's scheme structure. 

We consider the stack ${\mathbf P}^{n-1}(a_1,\ldots, a_n)$ from the point of view quotient stacks \cite{Auroux,Alberto}, where it was shown that it has a full and strong exceptional collection of length $n$:
\begin{equation}
\O,\; \O(1)\; \ldots\; \O(m-1)\,,\qquad \mbox{where $m=\sum_{i=1}^n a_i$}.
\end{equation}
The mutation-theoretic dual of this exceptional collection was throughly investigated in \cite{Alberto}.\footnote{For the definition and properties of mutations see, e.g., \cite{Rudakov:Book}.} 
In particular, Proposition 2.5.11 of \cite{Alberto} states that the mutation-theoretic left dual of the collection $\O, \ldots , \O(m-1)$  is given by the full exceptional sequence 
\begin{equation}
{\c M}_{(1-m)}\,,\; {\c M}_{(2-m)}\,,\; \ldots \,,\; {\c M}_{(-1)}\,,\; {\c M}_{(0)}\,.
\end{equation}

In order to explain the previous expression we need to introduce some notation. Let $I \subseteq \{1, 2,\ldots, n\}$ be a subset, and consider the  weighted projective stack ${\mathbf P}^{n-1}(a_1,\ldots, a_n)$. Then  $\# I$ will denote the number of elements in  $I$, while $a_I=\sum_{i\in I}a_i$. In this notation, for $0\geq l > -m $, the complex ${\c M}_{(l)}$ is defined as  a subcomplex of the Koszul complex $\c K$ twisted by $\O(l)$ \cite{Alberto}, with $j$th term given by:
\begin{equation}
{\c M}^j_{(-l)} = \bigoplus_{\#I=-j,a_I\leq l} \O(l -a_I ) \:\subseteq\: \bigoplus_{\#I=-j} \O(l -a_I ) = \c K^j(l)\,.
\end{equation}
In other words ${\c M}_{(l)}$ has non-zero components only in non-positive degrees.

Let us give two examples.
\begin{example}
For  $\mathbf{P}^1(a_1,a_2)$, assuming that $ a_1<a_2$,  the ${\c M}_{(l)}$'s are\footnote{We underlined the $0$th position in a complex.}
\begin{equation}\label{eq:albi12}
{\c M}_{(-l)} = \left\{ 
\begin{array}{rcl}
\xymatrix@1{\poso{\O(l)}} & 	 						       for  & 0\leq l<a_1 \\
\xymatrix@1{\O(l-a_1)\ar[r]& 		  \poso{\O(l)}} &	 		for  & a_1\leq l<a_2	\\
\xymatrix@1{\O(l-a_1)\oplus \O(l-a_2)\ar[r]&  \poso{\O(l)}}	  	&   for  & a_2\leq l <a_1+a_2	
\end{array}	\right.
\end{equation}
\end{example}

\begin{example}
For  $\mathbf{P}^1(1,1,3)$  the ${\c M}_{(l)}$'s are
\begin{equation}\label{eq:albi13}
\begin{split}
&{\c M}_{0} = \xymatrix@1{\poso{\O }}	\\
&{\c M}_{-1} = \xymatrix@1{\O^{\oplus 2}\ar[r]& 	 \poso{\O(1)}}	\\
&{\c M}_{-2} = \xymatrix@1{\O\ar[r]& \O(1)^{\oplus 2}\ar[r]& 	 \poso{\O(2)}}	\\
&{\c M}_{-3} = \xymatrix@1{\O(1)\ar[r]&  \O\oplus\O(2)^{\oplus 2}\ar[r]& 	 \poso{\O(3)}}	\\
&{\c M}_{-4} = \xymatrix@1{\O(1)^{\oplus 2}\oplus\O(2)\ar[r]&  \O(1)\oplus\O(3)^{\oplus 2} \ar[r]& 	 \poso{\O(4)}}	\\
\end{split}
\end{equation}
\end{example}

For brevity let us denote the stack $\mathbf{P}^{n-1}(a_1,\ldots, a_n)$ by ${\cal Y} $. Similarly, the partially resolved quotient stack $\mathrm{Bl} [\C^n/\Z_m]$, with exceptional divisor $D_v$, is denoted by  ${\cal X}$. Let 
$$i\!: {\cal Y}=\mathbf{P}^{n-1}(a_1,\ldots, a_n) \hookrightarrow  {\cal X}=\mathrm{Bl}[\C^n/\Z_m]$$
denote the embedding morphism of stacks. 
\begin{prop}
For any $n\geq 2$ and $\Z_m$ action on $\C^n$ such that the weights $(a_1,\ldots, a_n)$ do not have a common divisor and $m=\sum_{i=1}^n a_i$, the pushed-forward complexes 
$$i_*\c M_{(1-m)}\,,\;  i_*\c M_{(2-m)}\,,\; \ldots \,,\;  i_*\c M_{(-1)}\,,\;  i_*\c M_{(0)}$$ 
provide a model for the $\C^n/\Z_m$ fractional branes.
\end{prop}

\begin{proof}
The group $\Z_m$ has $m$ irreducible representations, all of them one dimensional. We label them by their characters: $\rho^i$, for $0\leq i <m$, where $\rho^m=1$. In terms of these the $\Z_m$ action (\ref{e:crepact}) is $\rho^{a_1}\oplus  \ldots \oplus \rho^{a_n}$. The McKay quiver has its nodes given by the irreps, and the number of arrows from $\rho^i$ to $\rho^j$ is given by $b_{i j}$ in the following formula
\begin{equation}\label{McKay}
\rho^i\otimes \left( \rho^{a_1}\oplus \ldots \oplus \rho^{a_n}\right) = \oplus_{j=1}^n b_{i j} \rho^j\,.
\end{equation} 
Therefore $b_{i j}=\# \{k\in \{1,\ldots,  n \}|\, a_k=j-i \}$, i.e., the cardinality of the finite set.

Now we fix a node, say $\rho^i$, and we analyze to what nodes do arrows go from this node. This information for all $i$ is sufficient to draw the quiver. By (\ref{McKay}) the arrows go to the nodes $\rho^{a_k+i}$. Now $0<a_k<m$ and $0\leq i<m$, and thus $0<a_k+i<2m$. In conclusion, there are arrows going from node $\rho^i$ to node $\rho^j$ if 
\begin{equation}\label{e:refMcK}
\begin{minipage}{5in}
\begin{itemize}
\item $i< j $ and $j=i+a_k $ for some $k$; or 
\item $i >j $ and $i+a_k=j+m$ for some $k$.
\end{itemize}	
\end{minipage}
\end{equation} 
Taking into account the multiplicities,  the number of arrows going  from node $\rho^i$ to node $\rho^j$ is
\begin{equation}\label{eQuivarr}
N_{i\to j}= \left\{ 
\begin{array}{ll}
\# \{k|\, a_k=j-i \} & \mbox{if $i< j$}	\\
\# \{k|\, a_k=m+j-i \}  & \mbox{if $i> j$}.
\end{array}	\right.
\end{equation} 

As we said, this is sufficient to draw the McKay quiver. Next we show that the same rules are satisfied by the $\Ext^1$ quiver of the proposed  fractional branes.

To evaluate the $\Ext$-groups we use the stacky version of the spectral sequence (\ref{SS1}) (see, e.g., \cite{Eric:DC2}). In our case $i\!: {\cal Y}= \mathbf{P}^{n-1}(a_1,\ldots, a_n)  \hookrightarrow  {\cal X}=\mathrm{Bl}_k [\C^n/\Z_m]$. For ${\cal E}$ and $ \cal F$ two objects in $\D({\cal Y})$ the spectral sequence reads
\begin{equation}\label{SS1a}
E_2^{p,q}=
\Ext^p_{\c Y}( {\cal E} ,\, {\cal F} \otimes \Lambda^q N_{\c Y/\c X}) \: \Longrightarrow \:
\Ext^{p+q}_{\c X}\left( i_* {\cal E}, i_* {\cal F} \right)\,.
\end{equation}
Since $N_{\c Y/\c X}=K_{\c Y}$ has rank one,\footnote{We used the fact that $K_{\c X}$ is trivial. } the spectral sequence degenerates at $E_2$, and we have that 
\begin{equation}
\Ext^1_{\c X}\left( i_* {\cal E}, i_* {\cal F} \right)=\Ext^1_{\c Y}( {\cal E} ,\, {\cal F})\oplus \Ext^0_{\c Y}( {\cal E} ,\, {\cal F}\otimes K_{\c Y}).
\end{equation}
Serre duality gives $\Ext^0_{\c Y}( {\cal E} ,\, {\cal F}\otimes K_{\c Y})=\Ext^{n-1}_{\c Y}( {\cal F} ,\, {\cal E})^\vee$, and therefore
\begin{equation}\label{SS1b}
\Ext^1_{\c X}\left( i_* {\cal E}, i_* {\cal F} \right)=\Ext^1_{\c Y}( {\cal E} ,\, {\cal F})\oplus \Ext^{n-1}_{\c Y}( {\cal F} ,\, {\cal E})^\vee\,.
\end{equation}

The  $\Ext^i_{\c Y}(\c M( i), \c M( j))$ groups are easily computed using Lemma~2.5.12 of \cite{Alberto}:
\begin{equation}
\dim \Ext^k_{\c Y}(\c M( i),\c M( j)) =  \#\{J \subseteq \{1, \ldots, n\} \, |\, \#J = k,a_J  = j -i\}\,.
\end{equation}
Therefore
\begin{equation}\label{eq1}
\dim \Ext^1_{\c Y}(\c M( i),\c M( j)) =  \#\{k \in \{1, \ldots, n\} \, |\, a_k  = j -i\}\,,
\end{equation}
and 
\begin{equation}\label{eq2}
\dim \Ext^{n-1}_{\c Y}(\c M( i),\c M( j)) =  \#\{k \in \{1, \ldots, n\} \, |\, m-a_k  = j -i\}\,,
\end{equation}
Observe that both (\ref{eq1}) and (\ref{eq2}) vanish if $i> j$, and thus only one of them contributes to (\ref{SS1b}) for any given $i$ and $j$. Putting the pieces together we have that 
\begin{equation}\label{SS1bb}
\Ext^1_{\c X}\left(i_*\c M( i),i_*\c M( j)   \right)=\left\{ 
\begin{array}{ll}
\# \{k|\, a_k=j-i \} & \mbox{if $i< j$}	\\
\# \{k|\,  m-a_k  = i-j \}  & \mbox{if $i> j$}.
\end{array}	\right.
\end{equation}
But this expression agrees with (\ref{eQuivarr}), and hence the two quivers are identical.

To complete the proof, we need to show that the $i_*\c M_{(l)}$'s  indeed ``add up'' to the D3-brane. It suffices to show the following K-theory relation
\begin{lemma}
\begin{equation}
\sum_{l=0}^{1-m} [  \c M_{(l)} ]= [ \O_{pt}]\,.
\end{equation} 
\end{lemma}

\begin{proof}
By definition
\begin{equation}
\sum_{l=0}^{m-1} [  \c M_{(-l)}] = \sum_{l=0}^{m-1} \sum_{j=0}^{n-1} (-1)^j\, [  \c M_{(-l)}^{-j}]=
\sum_{l=0}^{m-1}  \sum_{j=0}^{n-1} (-1)^j\, \sum_{\# I= j,\\ a_I\leq l} \, {(l-a_I)H}\,,
\end{equation} 
where we called $[ \O(1) ]= H$. The last expression is in fact a  sum over restricted  pairs $(l,I)$
\begin{equation}\label{eChern:id}
\sum_{(l,I):\, a_I\leq l}\, (-1)^{\# I}\, {(l-a_I)H}\,.
\end{equation} 
The idea of the proof is to reorganize the terms of this sum such that one has natural cancellations, and then recognize the remaining terms as having to do with a certain Koszul resolution. To this aim we first investigate the pairs $(l,I)$ from the above sum.

Since $I\subset \{1, \ldots, n\}$ its cardinality is $k$, for some $0\leq k \leq n$. These subsets fall into two categories: those that contain the element $1$ and those that don't. The first step is to show that the $I$'s containing $1$ cancel out many of the terms coming from the subset  $I- \{1\}$. 

Let $I$ contain $1$. The sum in (\ref{eChern:id}) runs over pairs $(l,I)$ such that $a_I\leq l<m$. The contribution of such a term is $(-1)^{\# I}\, {(l-a_I)H}$, with $0\leq l-a_I<m-a_I$, or 
\begin{equation}\label{echcan1}
(-1)^{\# I}\, {\rho\, H}\,,\qquad \mbox{for $0\leq \rho <m-a_I$}.
\end{equation}  

Now consider the subset $I^\circ=I- \{1\}$, assuming that $1\in I$, for which $a_{I^\circ}=a_I-a_1$. The range of $l$'s for a pair $(l,I^\circ)$ is $a_{I^\circ}\leq l<m$, i.e., $a_I-a_1\leq l<m$. The contributions coming from the $(l,I^\circ)$'s  is $(-1)^{\# I^\circ}\, {(l-a_{I^\circ})H}$, with $0\leq l-a_{I^\circ}<m-a_{I^\circ}=m-a_I+a_1$. Or using $\# I^\circ=\# I-1$:
\begin{equation}\label{echcan2}
-(-1)^{\# I}\, {\rho\, H}\,,\qquad \mbox{for $0\leq \rho <m-a_I+a_1$}.
\end{equation}   

Comparing (\ref{echcan1}) and (\ref{echcan2}) we see that  (\ref{echcan1}) cancels all the terms in (\ref{echcan2}), except  those in the range $m-a_I\leq \rho <m-a_I+a_1$.

Let us rephrase what we obtained. Given a pair $(l,I)$:
\begin{enumerate}
\item if $1\in I$ then $(l,I)$ gives no contribution to (\ref{eChern:id}) -- since it cancels part of the $(l-a_1,I- \{1\})$ contribution;
\item if $1\notin I$ then $(l,I)$ contributes to (\ref{eChern:id}) only if $m -a_1\leq l <m $ -- the others are canceled by  $(l+a_1,I\cup\{1\} )$.
\end{enumerate}

The second statement is justified so far only if $\# I <n-1$. So what happens for $\# I =n-1$? If $I'= \{2, \ldots, n\}$, then there is no contribution coming from $I\cup\{1\}= \{1, \ldots, n\}$, since the condition $l<m$ disallows any pair containing $\{1, \ldots, n\}$. On the other hand, $a_{I'}=\sum_{i=2}^{n}a_i=m-a_1$, and thus the condition $a_{I'}\leq l<m$ reads $m -a_1\leq l <m $, which is what we want.\footnote{There is no problem at the other end, where $I=\{1\}$, since $I'=I-\{1\}=\emptyset$, for which the condition $a_{I'}\leq l<m$ reads $0\leq l <m $, and no terms are missing for the assumed cancellation.}

Therefore the sum (\ref{eChern:id}) has been reduced to (by the above conclusion the sum is unrestricted)
\begin{equation}\label{eChern:id4}
\sum_{l=m -a_1}^{m-1}\sum_{1\notin I}\, (-1)^{\# I}\, {(l-a_I)H}=
\left( \sum_{l=m -a_1}^{m-1}{l}\right) \left( \sum_{1\notin I}\, (-1)^{\# I}\right)H
-a_1\left( \sum_{1\notin I}\, (-1)^{\# I}\, {a_I }\right)H .
\end{equation} 
The first sum is 
\begin{equation}
\sum_{1\notin I}\, (-1)^{\# I}=
\sum_{i=0}^{n-1} (-1)^i{n-1\choose i}=0.
\end{equation} 
The second factor comes from the Koszul complex of the point $[1,0,\ldots,0]$ on $\mathbf{P}^{n-1}(a_1,\ldots, a_n)$. The ideal in question is $(x_2,\ldots, x_n)$, and the $j$th term of the  Koszul complex is $\c K^j = \oplus_{\#J=j}\O(-a_J )$, where now $J\subset \{2, \ldots, n\}$. These $J$'s are precisely the $I$'s appearing in (\ref{eChern:id4}). Therefore
\begin{equation}\label{eChern:id5}
\sum_{1\notin I}\, (-1)^{\# I}\, {a_I H}= [ \O_{[1,0,\ldots,0]}]\,,
\end{equation} 
the K-class of the singular point $[1,0,\ldots,0]$. $a_1$ times this is a non-singular point on the stack $\mathbf{P}^{n-1}(a_1,\ldots, a_n)$, and this proves the lemma. 
\end{proof}
\end{proof}

\section*{Acknowledgments}

It is a pleasure to thank Tom Bridgeland, Alberto Canonaco, Emanuel Diaconescu, Mike Douglas, Alastair King, Tony Pantev and  Ronen Plesser for useful conversations.  I am especially indebted to Paul Aspinwall for enlightening discussions. 

\begin{appendix}

\section{Some useful spectral sequences}\label{s:ss} 

In the bulk of the paper we make extensive use of spectral sequences. We recall three of them here (for more details see \cite{en:fracC2}).

The simplest case concerns a {\em smooth} subvariety $S$ of a smooth variety $X$. Let $i\!: S \hookrightarrow X$ be the embedding, and $N_{S/X}$ the normal bundle of $S$ in $X$. Then for two locally free sheaves ${\cal E}$ and ${\cal F}$ on $S$, we have the first  spectral sequence:
\begin{equation}\label{SS1}
E_2^{p,q}=
\Ext^p_S( {\cal E} ,\, {\cal F} \otimes \Lambda^q N_{S/X}) \: \Longrightarrow \:
{\Ext}^{p+q}_X\left( i_* {\cal E}, i_* {\cal F} \right)
\end{equation}
where $\Lambda^q$ denotes the $q$th exterior power. 

A more general case is when you are given two nested embeddings: $j\!: T \hookrightarrow S$ and $i\!: S \hookrightarrow X$,  a  vector bundle  $\c F$ on $T$,  and a vector bundle $\c E$ on $S$. Then we have the spectral sequence:
\begin{equation}\label{SS2}
E_2^{p,q}= 
\Ext^p_T ({\cal E}|_T ,\,  {\cal F} \otimes \Lambda^q N_{S/X}|_T )
\: \Longrightarrow \: \Ext^{p+q}_X\left( i_* {\cal E}, j_* {\cal F} \right)
\end{equation}
The symbol $|_T$ means restriction to $T$.

The final and most general case deals with two subvarieties $T$ and $S$ of $X$. Now the embeddings are $i\!: S \hookrightarrow X$ and $j\!: T \hookrightarrow X$. Once again $\c F$ is a vector bundle on $T$,  and $\c E$ is a vector bundle  on   $S$. The spectral sequence is:
\begin{equation}\label{SS3}
E_2^{p,q} = \Ext^p_{S\cap T}({\cal E}|_{S \cap T},\,  
{\cal F}|_{S \cap T} \otimes \Lambda^{q-m} \tilde{N} \otimes \Lambda^{top} N_{S \cap T / T} )\Longrightarrow
{\Ext}^{p+q}_X \left( i_* {\cal E}, j_* {\cal F} \right)
\end{equation}
where $ \tilde{N}= TX|_{S \cap T}/(TS|_{S \cap T} \oplus TT|_{S \cap T})$ is a quotient of tangent bundles, while $m=\rk N_{S \cap T / T}$.

We also recall  a spectral sequence derived in Appendix~B of \cite{en:fracC2}. Let $X$ be a smooth algebraic variety. Consider two divisors  $C$ and $D$ on $X$, and the embedding maps: $i \! : C+D  \hookrightarrow X$ and $j \! : C \hookrightarrow X$. The divisor $C+D$ is reducible, and singular. For a coherent sheaf $\c F$ on $C$, the spectral sequence with the $E_2^{p,q}$ term
\begin{equation}\label{eq:dcx}
\begin{xy}
\xymatrix@C=4mm{
  \\
E_2^{p,q}\; = \\
\\	}
\end{xy}
\quad
\begin{xy}
\xymatrix@C=4mm{
  & 0& \\
   & \H^p (C,  \c F (C^2\!+\!CD))\\
   & \H^p (C,  \c F )}
\save="x"!LD+<-3mm,0pt>;"x"!RD+<0pt,0pt>**\dir{-}?>*\dir{>}\restore
\save="x"!LD+<0pt,-3mm>;"x"!LU+<0pt,-2mm>**\dir{-}?>*\dir{>}\restore
\save!RD+<0mm,-4mm>*{p}\restore
\save!LU+<-3mm,-5mm>*{q}\restore
\end{xy}
\end{equation}
converges to $\Ext_X^{p+q}(i_*\O_{C+D}, j_*\c F)$.

Although these spectral sequences were derived for sheaves, they extend to the derived category. It is also clear that (\ref{SS1}) is a particular case of (\ref{SS2}), which in turn is a particular case of (\ref{SS3}). 
\end{appendix}


\end{document}